\documentclass{article}
\usepackage{amsmath}
\usepackage{latexsym}
\usepackage{a4wide}
\usepackage{epsfig}

\newcommand{\ice}[1]{\relax}
\newcommand{\Lams}{\Lambda_{\overline{\rm MS}}}

\newcommand{\SU}{\mathrm{SU}}
\newcommand{\ms}{\mbox{{\sc ms}}} 
\newcommand{\msbar}{\overline{\mbox{{\sc ms}}}} 
\newcommand{\MOM}{\mbox{{\sc mom}}} 
\newcommand{\MOMt}{\widetilde{\mbox{{\sc mom}}}} 
 
\newcommand{\MOMg}{\widetilde{\mbox{{\sc mom}}}\mathrm{g}} 
\newcommand{\MOMgg}{\widetilde{\mbox{{\sc mom}}}\mathrm{gg}} 
\newcommand{\MOMq}{\widetilde{\mbox{{\sc mom}}}\mathrm{q}} 

\newcommand{\smsbar}{\overline{\mbox{{\scriptsize \sc ms}}}} 
 
\newcommand{\sMOMt}{\widetilde{\mbox{{\scriptsize \sc mom}}}} 
\newcommand{\sMOMh}{\widetilde{\mbox{{\scriptsize \sc mom}}}} 
\newcommand{\sMOMg}{\widetilde{\mbox{{\scriptsize \sc mom}}}\mathrm{g}} 
\newcommand{\sMOMgg}{\widetilde{\mbox{{\scriptsize \sc mom}}}\mathrm{gg}} 
\newcommand{\sMOMq}{\widetilde{\mbox{{\scriptsize \sc mom}}}\mathrm{q}} 

\newcommand{\ZP}{ZP} 
\newcommand{\SC}{\mathrm{R}} 
\newcommand{\bare}{\mathrm{B}}
\newcommand{\xil}{\xi_L}
\newcommand{\alvp}{h}
\newcommand{\Zalpha}{Z_h}
\newcommand{\ZSC}[1]{Z^{\SC}_{#1}(\alvp^{\SC},\mu,\epsilon)}

\newcommand{\Break}{ \nonumber \\ & & }

\catcode`\@=11
\def\slash{\mathpalette\make@slash}
\def\make@slash#1#2{\setbox\z@\hbox{$#1#2$}%
  \hbox to 0pt{\hss$#1/$\hss\kern-\wd0}\box0}
\catcode`\@=12 

\begin{document}

\title{
 \vskip-3cm{\baselineskip14pt
\centerline{\normalsize\hfill Freiburg-THEP 00/11}
  \centerline{\normalsize\hfill TTP00--15}
  \centerline{\normalsize\hfill hep-ph/0007088}
 }
 \vskip.7cm
 {\Large\bf{
Three-loop Three-Linear  Vertices and Four-Loop 
$\MOMt$ 
$\beta$ functions  
in massless QCD 
}
 \vspace{1.5cm}
 }
}

\author{{K.G. Chetyrkin}\thanks{Permanent address:
Institute for Nuclear Research, Russian Academy of Sciences,
60th October Anniversary Prospect 7a, Moscow 117312, Russia.} ${}^{a,b}$
\  and 
A. R\'etey${}^{b}$
  \\[3em]
 ${}^a${\it Fakult{\"a}t f{\"u}r Physik,}
\\
{\it Albert-Ludwigs-Universit{\"a}t Freiburg,
D-79104 Freiburg, Germany }
\\
${}^b${\it Institut f\"ur Theoretische Teilchenphysik,} \\
  {\it Universit\"at Karlsruhe, D-76128 Karlsruhe, Germany}
 }
\date{}
\maketitle

\begin{abstract}
\noindent In this paper we present a full set of 2- and 3-point functions
for massless QCD at three-loop order in the $\msbar$ scheme. The vertex
functions are evaluated at the asymmetric point with one vanishing
momentum. These results are used  to relate the $\msbar$
coupling constant to that of various momentum subtraction ($\MOM$)
renormalization schemes at three-loop order. With the help of the known
four-loop $\msbar$ $\beta$-function we  then  determine the four-loop
coefficients of the corresponding $\MOMt$ $\beta$-functions.

As an application we consider the momentum dependence (running) of the
three-gluon asymmetrical vertex recently computed within the lattice
approach by Ph. Boucaud et al.  in \cite{Boucaud:2000ey}. An account of
the four-loop term in the corresponding $\beta$-function leads to a
significant (around 30\%) decrease of the value of the
non-perturbative $1/p^2$ correction to the running found in
\cite{Boucaud:2000ey}.  

\end{abstract}

\setcounter{page}{1}
\newpage 

\section{Introduction}

Momentum subtraction schemes provide a possibility to define a
renormalization description for QCD in a regularization independent
way.  The concept of momentum subtraction is very old.
It played an important role in the discussion of renormalization
description dependence of physical quantities long ago
\cite{phrva:d20:1420,phrva:d24:1369,phrva:d26:2038,nupha:b203:472}.
Recently, the momentum subtraction approach has been heavily
used to relate lattice results for quark masses and coupling constants
to their perturbatively determined $\msbar$ counterparts
(see, e.g. 
\cite{hep-ph/9803491,hep-ph/9910332,Becirevic:1999kb,%
hep-lat/9510045,hep-lat/9605033,Alles:1997fa,hep-ph/9810437,%
hep-ph/9810322,hep-ph/9903364,hep-ph/9910204,hep-lat/9710044}).
In  \cite{Boucaud:2000ey,hep-ph/9810322,hep-ph/9903364} it has been argued that the
knowledge of three-loop coefficients for the corresponding
$\beta$-functions is necessary and even the four-loop contributions
should be taken in to account.  This is because the accessible energy
ranges in these calculations are just reaching a level where
perturbative QCD calculations start to be valid approximations.

A large subclass (in fact infinitely many) of $\MOM$ schemes can be
defined by subtracting vertices at the asymmetric point where one
external momentum vanishes.  We will call this point the zero point
(\ZP).  These schemes are referred to as $\MOMt$ schemes and were
introduced and discussed in some detail at two-loop order in
\cite{phrva:d24:1369}. A crucial fact for $\MOMt$ schemes is that
setting one external momentum to zero for all three 3-vertices of
massless QCD never will produce infrared divergencies.

In this paper we present the perturbative calculation of the gluon,
ghost and quark self-energies and  all  fundamental\footnote{That is appearing in the
QCD Lagrangian} 3-vertices  with one vanishing
external momentum for massless QCD in a general covariant gauge.
The one-loop triple gluon vertex was obtained in \cite{phrva:d22:2550}
in Feynman gauge, the result for general gauge can be found in
\cite{phrva:d54:4087}. At two loops, the triple gluon and ghost gluon
vertices at the \ZP\ have been determined in general gauge in
\cite{hep-ph/9801380}.  References to  earlier relevant publications   and
results for various  momentum configurations can also be found in this
work. The quark-gluon vertex can be found to two-loop order in
Feynman gauge in \cite{phrva:d24:1369}.

These three-loop results at hand allow one to relate the coupling
constants of any $\MOMt$-like scheme to the $\msbar$ scheme at
three-loop order.  Recently, in \cite{phlta:b400:379} even the
four-loop term of the $\msbar$ $\beta$-function has been computed.
Using this result, we can also determine the $\beta$-functions of any
such $\MOMt$-scheme up to (and including)  four loops.

The paper is organized as follows: In Section~\ref{sec:selfenergies}
the definitions of the ghost and quark self-energy and the gluon
polarization are introduced and a brief outline of their calculation
is given.  In Section~\ref{sec:vertices} we introduce the triple
gluon, quark gluon and ghost gluon vertices at the \ZP\ and our notation
for these along with a description of their calculation.  The
calculation of the triple gluon vertex is performed directly and
additionally in an independent way using the Ward-Slavnov-Taylor (WST)
identity relating the triple gluon vertex to the ghost gluon
vertex. This is content of Section~\ref{sec:WSTI}.  In
Section~\ref{sec:renmsb} the $\msbar$ renormalization procedure is
described.  In Sections~\ref{sec:momdefs} and \ref{sec:betamom} 
we   use the results of the previous Sections to
obtain  the coupling constants and $\beta$-functions for 4 different
$\MOMt$ schemes of particular interest from their $\msbar$
counterparts. 
Finally, in Section~\ref{discussion} we discuss  the numerical
importance of  our results on the example of a recent lattice
computation \cite{Boucaud:2000ey} of  the momentum dependence (running) of the
three-gluon asymmetrical vertex.

The complete results are given in the appendix and also
will be made available as input file for the algebraic programs {\sc
form} and {\sc mathematica} at:
\begin{center}
{\tt http://www-ttp.physik.uni-karlsruhe.de/Progdata/}
\end{center}

\section{The Gluon Polarization and the Ghost and Quark Self-Energies} 
\label{sec:selfenergies}

The QCD Lagrangian with $n_f$ massless quark flavors in the covariant
gauge is:
\begin{eqnarray}
{\cal L} & = & -\frac{1}{4} G^a_{\mu\nu} G^{a\mu\nu} 
+ \mathrm{i} \sum_{f=1}^{n_f} \bar \psi^f_i [\slash{D}]_{ij} \psi^f_j 
- \frac{1}{2\xi_L} (\partial^\mu A^a_\mu)^2 
+ \partial^\mu \bar\eta^a (\partial \eta^a - g f^{abc} \eta^b A^c_\mu)
{},
\\
G^a_{\mu\nu} & = & \partial_\mu A^a_\nu - \partial_\nu A^a_\mu +
g f^{abc} A^a_\mu A^b_\nu, \quad
[D_{\mu}]_{ij} = \delta_{ij}\partial_\mu - \mathrm{i} g A^a_\mu T^a_{ij}
{}.
\nonumber
\end{eqnarray}
The quark fields $\psi^f_i$ transform as the fundamental representation
and the gluon fields $A^a_\mu$ as the adjoint representation of the
gauge group $\SU(3)$. $T^a_{ij}$ and $f^{abc}$ are the generators of
the fundamental and adjoint representation of the corresponding Lie
algebra. The $\eta^a$ are the ghost fields and $\xil$ is the gauge
parameter ($\xil=0$ corresponds to  Landau gauge).

From this Lagrangian one can derive three types of 2-point functions
\begin{eqnarray}
\label{def:2point}
G^{(2)\,ab}_{\mu\nu}(q) & = & 
D^{ab}_{\mu\nu}(q) =
\mathrm{i}
\int \mathrm{d}x \, \mathrm{e}^{\mathrm{i} qx} 
\langle T [ A^a_\mu(x) A^b_\nu(0) ] \rangle 
\nonumber 
,\\
G^{(2)\,ab}(q) & = & 
\Delta^{ab}(q) =
\mathrm{i} 
\int \mathrm{d}x \, \mathrm{e}^{\mathrm{i} qx} 
\langle T [ \eta^a(x) \bar\eta^b(0) ] \rangle 
,\\
G^{(2)}_{ij}(q) & = & 
S_{ij}(q) = 
\mathrm{i} 
\int \mathrm{d}x \, \mathrm{e}^{\mathrm{i} qx} 
\langle T [ \psi_i(x) \bar\psi_j(0) ] \rangle 
\nonumber
{},
\end{eqnarray}
where as usual $T[A B]$ is the time ordered product of $A$ and $B$ and
from now on we will skip the flavor index of the quarks.  The
propagators in eq.~(\ref{def:2point}) can be expressed in terms of
the corresponding self-energies in the following way:
\begin{eqnarray}
\label{def:selfenergies}
D^{ab}_{\mu\nu}(q) & = & \frac{\delta^{ab}}{(-q^2)}\left[
(-g_{\mu\nu} + \frac{q_\mu q_\nu}{q^2}) \frac{1}{1+\Pi(q^2)} 
- \xil \frac{q_\mu q_\nu}{q^2}\right]
\nonumber {},\\
\Delta^{ab}(q) & = & \frac{\delta^{ab}}{(-q^2)} \frac{1}{1+\tilde\Pi(p^2)}
{},\\
S_{ij}(q) & = & \frac{\delta^{ij}}{(-q^2)} \frac{\slash{q}}{1+\Sigma_V(q^2)} 
{}.
\nonumber 
\end{eqnarray}
The self-energies $\Pi(q^2)$, $\tilde\Pi(q^2)$ and $\Sigma_V(q^2)$
can be calculated by applying the following projections to all one
particle irreducible (1PI) diagrams with two external legs of the
corresponding type:
\begin{eqnarray}
\label{proj:selfenergies}
\Pi(q^2) & = &  
\frac{\delta^{ab}}{N^2-1}
\left[ \frac{1}{D-1}(-g_{\mu\nu} + \frac{q_\mu q_\nu}{q^2})
\frac{1}{q^2} \right]
\times
\raisebox{-4.5ex}{\epsfig{file=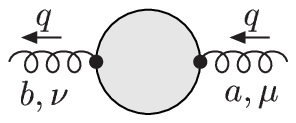}}
\nonumber {},\\
\tilde \Pi(q^2) & = &
\frac{\delta^{ab}}{N^2-1} \frac{1}{q^2}
\times
\raisebox{-4.5ex}{\epsfig{file=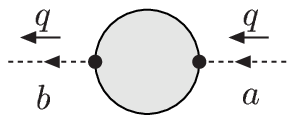}}
{},\\
\Sigma_V(q^2) & = & 
\frac{\delta^{ab}}{N} \mathrm{Tr} \left[
\frac{\slash{q}}{4q^2} 
\times 
\raisebox{-4.5ex}{\epsfig{file=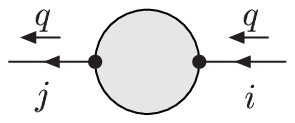}}
\right] 
\nonumber 
{}.
\end{eqnarray}
These projectors take into account that we are using dimensional
regularization with $D=4-2\epsilon$ and are valid for a general
$\SU(N)$ gauge group.  The trace has to be taken over Dirac matrices.
The resulting scalar integrals are of the massless propagator type and
can be evaluated with standard methods. For details about the
technical setup used to perform these calculations we refer to the
appendix.

\section{The Triple Gluon, Ghost and Quark Vertex Functions}
\label{sec:vertices}

In massless QCD there are the following 3-point functions:
\begin{eqnarray}
\label{def:3point}
G^{(3)\,abc}_{\mu\nu\rho}(p,q) & = & \mathrm{i}^2\int \mathrm{d}x \, \mathrm{d}y \,
\mathrm{e}^{-\mathrm{i}(px+qy)} 
\langle T [ A^a_\mu(x) A^b_\nu(y) A^c_\rho(0) ] \rangle
\nonumber {},\\
G^{(3)\,abc}_{\mu}(p,q) & = & \mathrm{i}^2\int \mathrm{d}x \, \mathrm{d}y \,
\mathrm{e}^{-\mathrm{i}(px+qy)} 
\langle T [ \eta^b(x) \bar\eta^c(y) A^a_\mu(0) ] \rangle
{},\\
G^{(3)\,a}_{\mu\,ij}(p,q) & = & \mathrm{i}^2\int \mathrm{d}x \, \mathrm{d}y \,
\mathrm{e}^{-\mathrm{i}(px+qy)} 
\langle T [ \psi_i(x) \bar\psi_j(y) A^a_\mu(0) ] \rangle
\nonumber
{}.\end{eqnarray}
These are related to the vertex functions by
\begin{eqnarray}
\label{def:vertices}
G^{(3)\,abc}_{\mu\nu\rho}(p,q) & = & 
D^{ad}_{\mu\mu'}(-p) 
D^{be}_{\nu\nu'}(-q) 
D^{cf}_{\rho\rho'}(-p-q) 
\Gamma^{def}_{\mu'\nu'\rho'}(p,q,-p-q) 
\nonumber {},\\
G^{(3)\,abc}_{\mu}(p,q) & = & 
\Delta^{ad}(-p)
\tilde \Gamma^{def}_{\mu'}(p,q;-p-q) 
\Delta^{eb}(q)
D^{cf}_{\mu'\mu}(q+p) 
{},\\
G^{(3)\,a}_{\mu\,ij}(p,q) & = & 
S_{ii'}(-p) 
\Lambda^{d}_{\mu'i'j'}(p,q;-q-p)
S_{j'j}(q) 
D^{ad}_{\mu'\mu}(p+q) 
\nonumber
{}.\end{eqnarray}
Setting one external momentum to zero leaves only one momentum which
can be used to construct tensor decompositions of the vertex
functions. The color and Lorentz structure of all vertices at the
\ZP\ and our conventions are  discussed in the next
subsections.

\subsection{The Triple Gluon Vertex}

For symmetry reasons setting any of the three external gluon momenta
of the triple gluon vertex to zero will lead to the  same
scalar functions. Up to two loops, the triple gluon vertex is known
to be proportional to the totally antisymmetric color structure
functions $f^{abc}$.  A color structure proportional to the totally
symmetric $d^{abc}$ will not have a divergent part.  In this work we
only consider the $f^{abc}$ part.

The triple gluon vertex needs to be totally symmetric under exchange of
any two of the (bosonic) gluons and we are interested in the part
proportional to $f^{abc}$. So the Lorentz structure for this part in
the case of one vanishing external momentum is limited to the
following three tensor structures, which are all antisymmetric under
exchange of the two gluons with non vanishing momentum:
\begin{eqnarray}
\label{form:gggdecomposition}
\Gamma^{abc}_{\mu\nu\rho}(q,-q,0) & = & -\mathrm{i} g f^{abc} 
\left(\rule{0ex}{3ex}
(2 g_{\mu\nu}q_\rho-g_{\mu\rho}q_\nu-g_{\rho\nu}q_\mu) T_1(q^2) 
\right. \nonumber \\ & & \left. 
-( g_{\mu\nu} - \frac{q_\mu q_\nu}{q^2}) q_\rho T_2(q^2) 
+ q_\mu q_\nu q_\rho T_3(q^2) \right)\quad .
\end{eqnarray}
Due to the WST identity  for the triple gluon vertex(for details see
Section~\ref{sec:WSTI}) the third function needs to vanish and finding
$T_3(q^2) = 0$ is a check for the correctness of the result.  The
functions $T_i(q^2)$ can directly be calculated by applying the
following projectors to the 1PI diagrams with 3 external gluon legs of
which one carries zero momentum:
\begin{eqnarray}
\label{form:gggprojectors}
P^{abc}_{1\,\mu\nu\rho}(q) & = & \frac{\mathrm{i}f^{abc}}{N(N^2-1)}\left[
\frac{1}{D-1} \left( \frac{q_\mu q_\nu q_\rho}{q^4} 
- \frac{g_{\nu\rho} q_\mu}{2 q^2} - \frac{g_{\mu\rho} q_\nu}{2 q^2} \right)
\right]
\nonumber {},\\
P^{abc}_{2\,\mu\nu\rho}(q) & = &  \frac{\mathrm{i}f^{abc}}{N (N^2-1)}\left[
\frac{1}{D-1} \left(\frac{3 q_\mu q_\nu q_\rho}{q^4} 
                         - g_{\mu\nu}\frac{q_\rho}{q^2} 
                         - g_{\mu\rho} \frac{q_\nu}{q^2}
                         - g_{\nu\rho} \frac{q_\mu}{q^2} \right)
 \right] 
\nonumber {},\\
P^{abc}_{3\,\mu\nu\rho}(q) & = &  \frac{\mathrm{i}f^{abc}}{N (N^2-1)}\left[
\frac{q_\mu q_\nu q_\rho}{q^6} \right]
{},
\end{eqnarray}
\begin{equation}
T_i(q^2) =  P_{i\,\mu\nu\rho}^{abc}(q) \times 
\raisebox{-8ex}{\epsfig{file=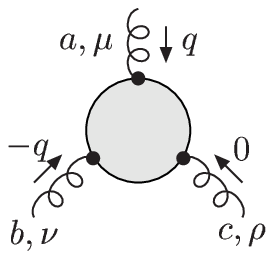}}
{}.
\end{equation}
The explicit calculation shows that indeed $T_3$ vanishes. 

\subsection{The Ghost Gluon Vertex}

The ghost gluon vertex is at tree level proportional to the momentum
of the outgoing ghost. Since this vertex is the only interaction of
the ghost field, it is clear that the vertex is also proportional to
this momentum at any order in perturbation theory. This leaves two
possibilities of one zero external momentum and due to the simple
Lorentz structure also only two scalar functions to be determined (again
we only consider only the $f^{abc}$ color structure):
\begin{eqnarray}
\tilde \Gamma^{abc}_{\mu}(-q,0;q) & = &
- \mathrm{i} g f^{abc} q_\mu \tilde \Gamma_{\mathrm{h}}(q^2)
\nonumber {},\\
\tilde \Gamma^{abc}_{\mu}(-q,q;0) & = &
- \mathrm{i} g f^{abc} q_\mu \tilde \Gamma_{\mathrm{g}}(q^2)
{}.\end{eqnarray}
The subscripts g and h stand for the external line that carries
zero momentum: g for the gluon and h for the ghost.
Again the $\tilde \Gamma_i(q^2)$ can be directly computed from the
1PI diagrams with two external ghosts and one external gluon:
\begin{eqnarray}
\tilde \Gamma_{\mathrm{h}}(q^2) & = &  
+ \frac{\mathrm{i} f^{abc}}{N (N^2-1)} \frac{q_\mu}{q^2} \times
\raisebox{-8ex}{\epsfig{file=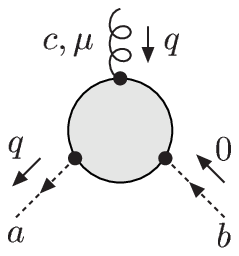}}
{},
\end{eqnarray}
\begin{eqnarray}
\tilde \Gamma_{\mathrm{g}}(q^2) & = & 
+ \frac{\mathrm{i} f^{abc}}{N (N^2-1)} \frac{q_\mu}{q^2} \times
\raisebox{-8ex}{\epsfig{file=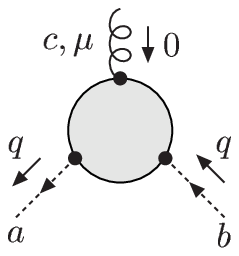}}
{}.
\end{eqnarray}

\subsection{The Quark Gluon Vertex}

Finally the quark gluon vertex is proportional to the color structure
$T^a_{ij}$ and for one vanishing external momentum there are two
different possibilities (here setting either of the quark momenta to
zero gives the same scalar functions up to three loops). A useful
tensor decomposition is:
\begin{eqnarray}
\Lambda^{a}_{\mu\,ij}(-q,0;q) & = &  g T^a_{ij} \left[
\gamma_\mu \Lambda_{\mathrm{q}}(q^2) +
\gamma_\nu \left( g_{\mu\nu} - \frac{q_\mu q_\nu}{q^2} \right) 
\Lambda_{\mathrm{q}}^T(q^2)
\right]
\nonumber 
{},
\\
\Lambda^{a}_{\mu\,ij}(-q,q;0) & = & 
g T^a_{ij} \left[
\gamma_\mu \Lambda_{\mathrm{g}}(q^2) +
\gamma_\nu \left( g_{\mu\nu} - \frac{q_\mu q_\nu}{q^2} \right) 
\Lambda_{\mathrm{g}}^T(q^2)
\right]
{},\end{eqnarray}
where the additional subscript q corresponds to the case of a
vanishing external quark momentum and $T$ marks the transversal part.
The $\Lambda_i$ can directly be calculated in the following way:
\begin{eqnarray}
\Lambda_{\mathrm{q}}(q^2) & = &  + \frac{T^a_{ij}}{N C_F}
Tr \left[ \frac{\slash{q} q_\mu}{4 q^2} \times 
\raisebox{-8ex}{\epsfig{file=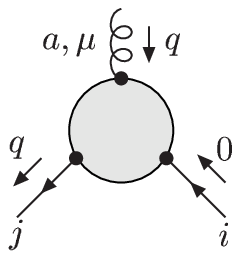}}
\right] \nonumber
{},
\end{eqnarray}
\begin{eqnarray}
\Lambda^T_{\mathrm{q}}(q^2)  & = & \frac{T^a_{ij}}{N C_F}
Tr \left[ \frac{1}{4(D-1)} \left( 
\gamma_\mu - D \frac{\slash{q} q_\mu}{q^2} \right) \times 
\raisebox{-8ex}{\epsfig{file=qqg_q.eps}}
\right] 
{},
\end{eqnarray}
\begin{eqnarray}
\Lambda_{\mathrm{g}}(q^2)  & = & 
 + \frac{T^a_{ij}}{N C_F}
Tr \left[ \frac{\slash{q} q_\mu}{4 q^2} \times 
\raisebox{-8ex}{\epsfig{file=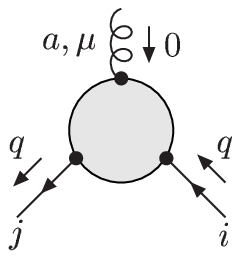}}
\right] 
{},
\nonumber 
\end{eqnarray}
\begin{eqnarray}
\Lambda^T_{\mathrm{g}}(q^2)  & = & \frac{T^a_{ij}}{N C_F}
Tr \left[ \frac{1}{4(D-1)} \left(
\gamma_\mu - D \frac{\slash{q} q_\mu}{q^2} \right) \times 
\raisebox{-8ex}{\epsfig{file=qqg_g.eps}}
\right] 
{}.
\end{eqnarray}
\section{The Ward-Slavnov-Taylor Identity}
\label{sec:WSTI}        

By using the WST identity for the triple gluon vertex, one can
determine the scalar functions $T_1$ and $T_2$ from the gluon and
ghost self-energies and functions related to the ghost gluon
vertex. This check has also been performed in other determinations of
the gluon vertex \cite{phrva:d22:2550} \cite{phrva:d54:4087}
\cite{hep-ph/9801380}.  The general WST identity was determined in
\cite{tmpha:10:99} \cite{nupha:b33:436} and can be found in the
following form for the triple gluon vertex in \cite{phrva:d22:2550}
\cite{phrva:d54:4087}:
\begin{eqnarray}
\label{form:stifull}
k^{\rho} \Gamma_{\mu\nu\rho}(p,q,k) & = &
- J(p^2) G(k^2) ( g^{\;\rho}_\mu p^2 - p^\rho p_\mu ) 
\tilde \Gamma_{\rho\nu}(p,k;q) 
\nonumber \\ & &
+ J(q^2) G(k^2) ( g^{\;\rho}_\nu q^2 - q^\rho q_\nu ) 
\tilde \Gamma_{\rho\mu}(q,k;p)
{}.
\end{eqnarray}
Here $\tilde \Gamma_{\nu\mu}(p,q;k)$ is related to the proper
ghost gluon vertex $\tilde \Gamma^{abc}_{\mu}(p,q;k)$ by: 
\begin{equation}
\label{def:ghostvertexmunu}
\tilde \Gamma^{abc}_{\mu}(p,q;k) = - \mathrm{i} g f^{abc} p^\nu 
\tilde \Gamma_{\nu\mu}(p,q;k) 
{}.
\end{equation}
and we have introduced the functions
\begin{equation}
J(p^2) = 1+\Pi(p^2) \quad \mathrm{and} \quad 
G(p^2) = \frac{1}{1+\tilde\Pi(p^2)} 
{} \ . 
\nonumber 
\end{equation}
In ref.~\cite{phrva:d22:2550} the following tensor decomposition for
$\tilde \Gamma_{\mu\nu}(p,q;k)$ is given:
\begin{eqnarray}
\label{form:gccdecomposition}
\tilde \Gamma_{\mu\nu}(p,k;q) & = &
+ g_{\mu\nu} a(q,k,p) 
- q_\mu k_\nu b(q,k,p)
+ p_\mu q_\nu c(q,k,p) 
\nonumber \\ & &
+ q_\mu p_\nu d(q,k,p)
+ p_\mu p_\nu e(q,k,p)
{}.
\end{eqnarray}
The analytic determination of the full momentum dependence of the
generalized ghost gluon vertex $\tilde \Gamma_{\nu\mu}$ at three loops
is impossible with the current calculation techniques. But of course
we do not need to know the full momentum dependence for checking the
WST identity at the \ZP. Still, even the determination of all possible
tensor structures for the case of one vanishing momentum and the
corresponding expansions to first order in the vanishing momentum
would be a very demanding project. To avoid the calculation of
unnecessary parts of the ghost gluon vertex we follow closely the
approach of \cite{hep-ph/9801380}.

For the vertex function there are two independent momenta. They can be
chosen in such a way that at the \ZP\ one of them
vanishes. Contracting the vertex in the WST identity
(\ref{form:stifull}) with the momentum that does not vanish at the
\ZP\ is relatively simple. In this case one can safely set
the other momentum to zero on the right hand side and is left
with functions of just one momentum. Inserting the tensor
decompositions (\ref{form:gggdecomposition}) and
(\ref{form:gccdecomposition}) gives:
\begin{equation}
\label{form:stiT1}
T_1(p^2) = a_3(p^2) G(p^2 ) J(p^2)\quad,\quad a_3(p^2) = a(0,p,-p) 
{}.
\end{equation}
There is also the possibility to contract the vertex with the momentum
that vanishes in the limit of the \ZP .  This case gives a
differential WST identity at the \ZP .  If the right hand
side is expanded to first order in $k_\rho$ the constant lowest order
term cancels. Dividing by $k_\rho$, setting $k$ to $0$ on both sides
and taking into account that for {\it massless} quarks $G(0) =
1$\footnote{$G(0)=1$ since all contributing diagrams are massless
tadpoles which vanish {in dimensional regularization.}} results in the
following representation of this differential identity:
\begin{eqnarray}
\label{form:stidiff}
\Gamma_{\mu\nu\rho}(p,-p,0) & = &
 + a_2(p^2) J(p^2) 
     \left( 2 g_{\mu\nu} p_\rho - g_{\mu\rho} p_\nu - g_{\nu\rho} p_\mu \right)
\nonumber \\ & &
 + a_2(p^2) p^2 \frac{\mathrm{d} J(p^2)}{\mathrm{d} p^2} 
      \left( - 2 \frac{p_\mu p_\nu p_\rho}{p^2} + 2 g_{\mu\nu} p_\rho \right)
\nonumber \\ & &
 + d_2(p^2) p^2 J(p^2) \left( 
     - 2 \frac{p_\mu p_\nu p_\rho}{p^2} + g_{\mu\rho} p_\nu + g_{\nu\rho} p_\mu
      \right)
\nonumber \\ & &
 + J(p^2) \left( a'_{23}(p^2) - a'_{21}(p^2) \right)
 \left( \frac{p_\mu p_\nu p_\rho}{p^2} - g_{\mu\nu} p_\rho \right)
{}.
\end{eqnarray}
where we have expanded $a(p,k,q)$ to first order in small $k$ and use
the following shortcuts:
\begin{eqnarray}
d(p,0,-p) = d_2(p^2) & , & a(p,0,-p) = a(-p,0,p) = a_2(p^2) 
\nonumber 
{},
\\
a(-p-k,k,p) & = & a_2(p^2) + \frac{k \cdot p}{p^2} a'_{23}(p^2) 
+ {\cal O}(k^2)\; ,
\\ 
a(p,k,-p-k) & = & a_2(p^2) +\frac{k \cdot p}{p^2} a'_{21}(p^2) 
+ {\cal O}(k^2) \; .\nonumber
\end{eqnarray}
Contracting eq.~(\ref{form:stidiff}) with $p^\mu$ once more and
projecting out the structure $T_1(p^2)$ gives another possibility to
reproduce $T_1(p^2)$. It can be used to produce the simple relation
\begin{equation} 
a_2(p^2) - p^2 d_2(p^2) = G(p^2) a_3(p^2)
{}.
\end{equation}
Using this relation and projecting out the structure proportional to
$T_2(p^2)$ we can also obtain $T_2(p^2)$ in a completely
independent way:
\begin{eqnarray}
\label{form:stiT2}
T_2(p^2) & = & 2 T_1(p^2) 
        - 2 a_(p^2) \frac{\mathrm{d}}{\mathrm{d} p^2} \left(p^2 J(p^2)\right)
\nonumber \\ & &
        + J(p^2) \left( a'_{23}(p^2) - a'_{21}(p^2) \right)
{}.
\end{eqnarray}
Finally, applying the projector $P_3$ of
eq.~(\ref{form:gggprojectors}) to the right hand side of
eq.~(\ref{form:stidiff}) should give an alternative representation of
$T_3(p^2)$ and it is indeed found to be zero from this identity, as
has been mentioned in the last Section.

The scalar functions $a_2$, $a_3$, $a'_{23}$ and $a'_{21}$ can be
calculated from 1PI diagrams with the following operations:
\begin{equation}
P^{abc}_{\mu\nu}(q) = \frac{f^{abc}}{N (N^2-1)}
\frac{1}{D-1} \left( q^2 g_{\mu\nu} - q_\mu q_\nu \right)
{},
\end{equation}
\begin{equation}
a_2(q^2) = 1 + P^{abc}_{\mu\nu}(q) \times \left[
\raisebox{-8ex}{\epsfig{file=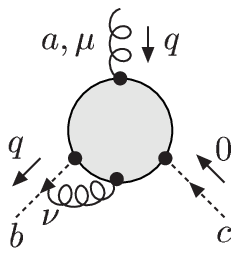}}
\right]
{},
\end{equation}
\begin{equation}
a_3(q^2) = 1 + P^{abc}_{\mu\nu}(q) \times \left[
\raisebox{-8ex}{\epsfig{file=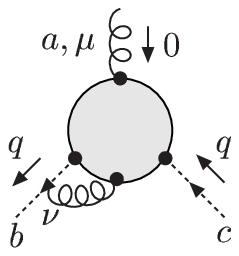}}
\right]
{},
\end{equation}
\begin{equation}
a'_{23}(q^2) = \Box_k \left( q \cdot k \, P^{abc}_{\mu\nu}(q) 
{\cal T}^{(1)}_k \times \left[
\raisebox{-8ex}{\epsfig{file=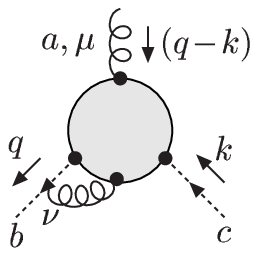}}
\right]\right)
{},
\end{equation}
\begin{equation}
a'_{21}(q^2) = \Box_k \left( q \cdot k \, P^{abc}_{\mu\nu}(q) 
{\cal T}^{(1)}_k \times \left[
\raisebox{-8ex}{\epsfig{file=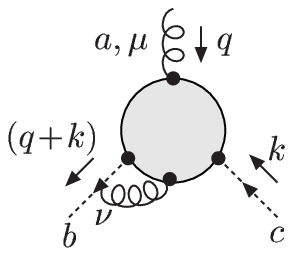}}
\right]\right)
{}.
\end{equation}
The additional gluon line is to visualize that at the outgoing ghost
vertex we do not contract with the external momentum, which leaves an
additional open index $\nu$ as shown in
eq.~(\ref{def:ghostvertexmunu}). The shortcut ${\cal T}^{(1)}_{k}$
stands for a first order Taylor expansion in $k$ and $\Box_k =
g_{\alpha\beta}\frac{\partial}{\partial k_\alpha}
\frac{\partial}{\partial k_\beta}$. Both are to be applied to the
integrand {\bf before} any of the integrations are performed. These
manipulations are necessary because the {\sc mincer} package can only
handle scalar products of internal and one external momentum for
more than one loop. Since there are only the two independent external
momenta $p$ and $k$, the first order terms in $k$ (of a scalar
function) have to be proportional to $k \cdot p$. This guarantees that
the above manipulations will extract the scalars $a'_{2i}$.

Using this strategy we found complete agreement for the bare
expression with the ones from the direct computation of the triple
gluon vertex.

\section{Renormalization}
\label{sec:renmsb}

For a generic renormalization scheme\footnote{In the following all
quantities without explicit superscript refer to the $\msbar$ scheme,
the superscript $\bare$ marks the bare quantities} $\SC$, the
following relations between bare and renormalized quantities hold in
dimensional regularization:
\begin{eqnarray}
\label{def:ren}
(A^{\bare})^{a}_{\nu} = \sqrt{\ZSC{3}} \, (A^{\SC})^{a}_{\nu}(\mu)\;,
& & 
\eta^{\bare} = \sqrt{\tilde \ZSC{3}} \, \eta^{\SC}(\mu)\;,
\nonumber \\ 
\psi_i^{\bare} = \sqrt{\ZSC{2}} \, \psi_i^{\SC}(\mu),
& &
\alvp^{\bare} = \mu^{2\epsilon} \, \ZSC{h} \, \alvp^{\SC}(\mu)
{}.
\end{eqnarray}
Here $\alvp = \alpha_s/(4\pi) = g_s^2/(16\pi^2)$, $\mu$ is the {}'t Hooft
unit of mass which is the renormalization point in the $\ms$
scheme. 

The self-energies and vertex functions are renormalized as follows 
\begin{eqnarray}
\label{ren:green}
1+\Pi^{\SC}(q^2,\alvp^R,\mu,\epsilon) & = & 
\ZSC{3} \left[ 1+ \Pi^{\bare}(q^2,\alvp^{\bare},\epsilon) 
\right]_{\alvp^{\bare} = \mu^{2\epsilon} \Zalpha \alvp,\,\xil^{\bare}=Z_3\xil}
\nonumber {},\\
1+\tilde\Pi^{\SC}(q^2,\alvp^R,\mu,\epsilon) & = &
\tilde \ZSC{3} \left[ 1+\tilde\Pi^{\bare}(q^2,\alvp^{\bare},\epsilon) 
\right]_{\alvp^{\bare} = \mu^{2\epsilon} \Zalpha \alvp,\,\xil^{\bare}=Z_3\xil}
\nonumber {},\\
1+\Sigma_V^{\SC}(q^2,\alvp^R,\mu,\epsilon) & = & 
\ZSC{2} \left[ 1+\Sigma_V^{\bare}(q^2,\alvp^{\bare},\epsilon) 
\right]_{\alvp^{\bare} = \mu^{2\epsilon} \Zalpha \alvp,\,\xil^{\bare}=Z_3\xil}
\nonumber {},\\
T^{\SC}_{1,2}(q^2,\alvp^R,\mu,\epsilon) & = & \left.
\ZSC{1} T^{\bare}_{1,2}(q^2,\alvp^{\bare},\epsilon) 
\right|_{\alvp^{\bare} = \mu^{2\epsilon} \Zalpha \alvp,\,\xil^{\bare}=Z_3\xil}
\nonumber {},\\
\tilde \Gamma^{\SC}_{h,g}(q^2,\alvp^R,\mu,\epsilon) & = & \left.
\tilde \ZSC{1} \tilde \Gamma^{\bare}_{h,g}(q^2,\alvp^{\bare},\epsilon) 
\right|_{\alvp^{\bare} = \mu^{2\epsilon} \Zalpha \alvp,\,\xil^{\bare}=Z_3\xil}
\nonumber {},\\
\Lambda^{\SC}_{g,q}(q^2,\alvp^R,\mu,\epsilon) & = & \left.
\bar \ZSC{1} \Lambda^{\bare}_{g,q}(q^2,\alvp^{\bare},\epsilon) 
\right|_{\alvp^{\bare} = \mu^{2\epsilon} \Zalpha \alvp,\,\xil^{\bare}=Z_3\xil}
{}.
\end{eqnarray}
\ice{
where the gauge parameter ``renormalization'' takes into account that
the gauge fixing term in the QCD Lagrangian ${\cal L}_{\mathrm{gf}}=
1/\xil \, \partial^\mu A^a_\mu \partial^\nu A^a_\nu$ does not get
counterterms.  
}
The renormalization constants for the self-energies can
be read of eqs.(\ref{def:2point},\ref{def:selfenergies}).  Likewise,
from eqs.~(\ref{def:3point},\ref{def:vertices}) and the QCD Lagrangian
the renormalization factors for the trilinear operators can be related
to the renormalization of the coupling constant and the field
renormalizations
\begin{equation}
\label{Zrelations}
\sqrt{Z^R_a Z^R_3} = \frac{Z^R_1}{Z^R_3} 
= \frac{\tilde Z^R_1}{\tilde Z^R_3} = \frac{\bar Z^R_1}{Z^R_2} \,.
\end{equation}
QCD is known to be a renormalizable theory. This means that
$Z^R_x$ can be found which make the renormalized Greens functions
eqs.~(\ref{def:2point},\ref{def:3point}) finite when taking the limit
$\epsilon \to 0$ and at the same time fulfill the WST identities,
respectively eq.~(\ref{Zrelations}), to any order in perturbation
theory.

Using all these formulas, renormalization in the $\msbar$-scheme is
straightforward. It requires the $Z_x=Z_x^{\smsbar}(\alvp,\epsilon)$
to only contain poles in $\epsilon$ and thus to be of the following
form:
\begin{equation}
Z_x(\alvp,\epsilon)= 1 + \sum_{i>0} \frac{1}{\epsilon^i} Z^{(i)}_x(\alvp)
\quad,\quad
Z_x^{(i)}(\alvp)= 1 + \sum_{j \ge i} \alvp^j Z^{(i,j)}_x
\end{equation}
The $\mu$-dependence of the fields and the coupling $\alvp$ is then
described in the usual way by the renormalization group equations:
\begin{eqnarray}
\mu^2 \frac{\mathrm{d}}{\mathrm{d}\mu^2} \alvp(\mu) = -\epsilon \alvp + \beta(\alvp) 
&,&
\beta(\alvp) = - \sum_{i=0}^{\infty} \alvp^{(i+2)}\beta_{i} = 
\alvp^2 \frac{\partial}{\partial \alvp} \Zalpha^{(1)} 
\nonumber {},\\
2 \mu^2 \frac{\mathrm{d}}{\mathrm{d}\mu^2} A^a_\nu(\mu) =  
\gamma_{3}(\alvp) A^a_\nu(\mu) 
&,&
\gamma_{3}(\alvp) = - \sum_{i=0}^{\infty} \alvp^{(i+1)}\gamma_{3\,i}= 
\alvp \frac{\partial}{\partial \alvp} Z_3^{(1)} 
\nonumber {},\\
2 \mu^2 \frac{\mathrm{d}}{\mathrm{d}\mu^2} \eta(\mu) =  
\tilde \gamma_{3}(\alvp) \eta(\mu) 
&,&
\tilde \gamma_{3}(\alvp) = - \sum_{i=0}^{\infty} \alvp^{(i+1)} \tilde
\gamma_{3\,i} = \alvp \frac{\partial}{\partial \alvp} \tilde Z_3^{(1)} 
\nonumber {},\\
2 \mu^2 \frac{\mathrm{d}}{\mathrm{d}\mu^2} \psi_i(\mu)  =  
\gamma_{2}(\alvp) \psi_i(\mu)
&,&
\gamma_2(\alvp) = - \sum_{i=0}^{\infty} \alvp^{(i+1)}\gamma_{2\,i} = 
\alvp \frac{\partial}{\partial \alvp} Z_2^{(1)} 
{},
\end{eqnarray}
where the last equalities can be derived from
eqs.~(\ref{def:ren}), the $D$ dimensional $\beta$-function
$[-\epsilon \alvp + \beta(\alvp)]$ and the fact that the bare quantities
must be independent of $\mu$. As the functions in
eq.~(\ref{ren:green}) $\beta$ and the anomalous dimensions are here
defined in $D$ dimensions but are finite for $\epsilon \to 0$.

The $\msbar$ renormalization constants and anomalous dimensions are
well known up to three loops 
\cite{Tar82,tar}
 and for the $\beta$-function
\cite{phlta:b400:379} and the quark field anomalous dimension
$\gamma_2$ \cite{hep-ph/9910332} even the four-loop terms are known.
Of course up to three loops the $Z_i$ can also be determined
independently from the bare results for the self-energies and vertex
functions.  The corresponding anomalous dimensions are given in the
appendix.

Performing this standard renormalization procedure one arrives at the
$\msbar$ renormalized expressions for all self-energies and vertex
functions.  The expressions for these in the limit $\epsilon \to 0$ at
the point $p^2=-\mu^2$ are given in the appendix for generic color
factors and a generic gauge parameter. The momentum dependence can be
obtained from the anomalous dimensions.  The fact that indeed all
poles in $\epsilon$ cancel and eq.~(\ref{Zrelations}) holds is another
check for the two-loop finite part and at least for  the pole terms of any of
the three-loop results.

\ice{
\begin{center} ?????????????????????????????????????????? \end{center} 
It is interesting to note that the scalar functions $a_3$ and $a_2$
appearing in the ghost gluon vertex can also be renormalized with
$\tilde Z_1$ as is expected. On the other hand, only the combination
$a^{\prime}_{23}+a^{\prime}_{21}$ can be renormalized this way, since
both show poles even at one-loop order although they do not appear at
tree level.  In the combination $a^{\prime}_{23}+a^{\prime}_{21}$
these poles cancel at one loop and the poles at higher orders can be
renormalized by $\tilde Z_1$.  It is remarkable that this is {\it not}
the combination that appears in the WST identity.  Power counting
indicates that these could be infrared poles. All other functions in
the WST identity are finite after renormalization. This means that for
some reason the renormalized functions do not fulfill the WST identity
at the \ZP ???
\begin{center} ??????????????????????????????????????????? \end{center} 
 }

\section{Four Particular $\MOMt$ like Scheme Definitions}
\label{sec:momdefs}

Momentum subtraction schemes are defined by setting some of the 2 and
3-point functions to their tree values for a fixed configuration of
the states of the external particles (that is momenta and polarization
state) in a certain gauge. This fixes the field renormalization
constants at the renormalization point $\mu$:
\begin{eqnarray}
\label{subtract:props}
1+\Pi^{\sMOMt}(-\mu^2) =  1 & = & 
Z_3^{\sMOMt}(\mu^2,\epsilon) \left[ 1+\Pi^{\bare}(-\mu^2,\epsilon) \right]
\nonumber {},\\
\left[1+\tilde\Pi^{\mathrm{\sMOMt}}(-\mu^2) \right] = 1 & = &
\tilde Z_3^{\sMOMt}(\mu^2,\epsilon) 
\left[ 1 + \tilde\Pi^{\bare}(-\mu^2,\epsilon) \right]
\nonumber {},\\
\left[ 1+\Sigma_V^{\sMOMt}(-\mu^2,\epsilon) \right]
 = 1 & = & Z_2^{\sMOMt}(\mu^2,\epsilon) 
\left[ 1+\Sigma_V^{\bare}(-\mu^2,\epsilon) \right]
{}.
\end{eqnarray}
For the renormalization of the coupling constant there are infinitely
many possibilities to define a momentum subtraction renormalization
scheme, even when considering only the \ZP . Not only is
there an ambiguity in which vertex to subtract, but also there is the
freedom to use  a
certain linear combination of the scalar functions appearing in the
gluon and quark vertices, which can be related to fixed polarization
states of the external particles~\cite{phrva:d24:1369}.  If one considers the
generalized ghost gluon vertex as defined in
eq.~(\ref{def:ghostvertexmunu}) there are even more possibilities.

Each of these choices defines a different renormalization scheme just
as a different choice of the arbitrary renormalization scale.  In
general it is not possible to fix more than one vertex to it's tree
value at the \ZP\ with $p^2=-\mu^2$, since they all are related by
Ward identities.

``The'' $\MOMt$ scheme originally was defined in~\cite{phrva:d24:1369} as a
scheme in which all three triple-vertices are subtracted at the
\ZP\ without violating the Ward identities. This can be
achieved by choosing a special set of the appearing scalar functions at
one-loop order. Of course this universality does not hold anymore when
considering the two- and three-loop order. This is why even in the original
publication a generalization of this scheme for two and more loops is
given. It subtracts the ghost vertex and corresponds to the original
definition at one-loop order. Here we also consider the other two
schemes that coincide with the $\MOMt$ scheme at one-loop order. They
are defined by subtracting just the scalar functions that appear at
tree level of the triple gluon (we will refer to it as the $\MOMg$
scheme) and quark gluon vertex (quark momentum set to zero, $\MOMq$),
respectively.

Another $\MOMt$ like renormalization scheme definition ($\MOMgg$) that
uses the gluon vertex lately has been used to relate lattice results
of the triple gluon vertex~\cite{hep-ph/9810437} and the gluon
propagator~\cite{hep-ph/9903364} to perturbative calculations.  The
$\MOMq$ scheme that subtracts the quark gluon vertex has also been
used in lattice calculations~\cite{hep-lat/9710044}, but is not used
beyond one-loop order in this work.

\subsection{Subtracting the Ghost Gluon Vertex}

A generalization of the original $\MOMt$ scheme to more than one loop is
given by subtracting the ghost gluon vertex with the moment of the
incoming ghost set to zero and a longitudinal polarization of the
external gluon (the case of a transversally polarized
gluon will vanish at the \ZP).  The renormalization
condition is:
\begin{equation}
\tilde \Gamma_h^{\sMOMh}(-\mu^2) 
=
1 = \tilde Z^{\sMOMh}_1(\mu^2) \tilde \Gamma^{\bare}(-\mu^2)
{}.
\end{equation}
Using eqs.~(\ref{def:ren}-\ref{Zrelations},\ref{subtract:props}), we
can connect the coupling constant in this scheme to quantities
calculable in the $\msbar$ scheme through the following chain of
equations:
\begin{eqnarray}
g_s^{\sMOMh}(\mu) & = & \mu^{-\epsilon} g_s^b \frac{1}{\tilde
Z_g^{\sMOMh}(\mu)} = \mu^{-\epsilon} g_s^b \frac{\tilde
Z_3^{\sMOMh}(\mu) \sqrt{Z_3^{\sMOMh}(\mu)}} {\tilde Z_1^{\sMOMh}(\mu)}
\nonumber \\ & = & g_s^{\smsbar}(\mu) Z_g^{\smsbar}(\mu)\,\left[
\frac{\tilde \Gamma_h^{\bare}(p^2)}{(1+\tilde\Pi^{\bare}(p^2)) 
\sqrt{1+\Pi^{\bare}(p^2)}}
\right]_{p^2=-\mu^2} \nonumber\\
& = & g_s^{\smsbar}(\mu) \,\left[
\frac{\tilde \Gamma_h^{\smsbar}(p^2)}{
(1+\tilde\Pi^{\smsbar}(p^2)) 
\sqrt{1+\Pi^{\smsbar}(p^2)}} \right]_{p^2=-\mu^2}
{}.
\end{eqnarray}
Squaring this and inserting the $\msbar$ expressions for $\Pi$,
$\tilde\Pi$ and $\tilde \Gamma_h$ we arrive at the following relation
between the coupling in this scheme and the $\msbar$ coupling,
expressed as an expansion in $\alvp=\alvp^{\smsbar}$ and\footnote{see
Section~\ref{sec:betamom} for details} $\xil=\xil^{\smsbar}$:

\begin{eqnarray}
\label{amomhQCD}
\alvp^{\sMOMh} & = & 
 \alvp\;
+ \;     \alvp^{2}\;
\left[
    + \frac{169}{12} 
    - \frac{10}{9}      n_f
    + \frac{9}{2}      \xil
    + \frac{3}{4}      \xil^{2}
\right] 
+ \;     \alvp^{3}\;
\left[
    + \frac{76063}{144} 
    - \frac{4}{3}      n_f     \zeta_3
    - 5     n_f     \xil
\right. \Break \left. \qquad \qquad
    - \frac{1913}{27}      n_f
    + \frac{100}{81}      n_f^{2}
    + \frac{117}{8}      \zeta_3     \xil
    - \frac{351}{8}      \zeta_3
    + \frac{1719}{16}      \xil
    + \frac{549}{16}      \xil^{2}
    + \frac{81}{16}      \xil^{3}
\right]\Break
+ \;     \alvp^{4}\;
\left[
    + \frac{42074947}{1728} 
    - 159     n_f     \zeta_3     \xil
    + \frac{3}{4}      n_f     \zeta_3     \xil^{2}
    + \frac{8362}{27}      n_f     \zeta_3
    + \frac{2320}{9}      n_f     \zeta_5
    - \frac{6931}{16}      n_f     \xil
\right. \Break \left. \qquad \qquad
    - \frac{757}{16}      n_f     \xil^{2}
    - \frac{769387}{162}      n_f
    + \frac{16}{3}      n_f^{2}     \zeta_3     \xil
    + \frac{28}{9}      n_f^{2}     \zeta_3
    + \frac{38}{9}      n_f^{2}     \xil
    + \frac{199903}{972}      n_f^{2}
\right. \Break \left. \qquad \qquad
    - \frac{1000}{729}      n_f^{3}
    + \frac{4893}{4}      \zeta_3     \xil
    + \frac{1485}{16}      \zeta_3     \xil^{2}
    - \frac{1341}{32}      \zeta_3     \xil^{3}
    - \frac{117}{32}      \zeta_3     \xil^{4}
    - \frac{60675}{16}      \zeta_3
\right. \Break \left. \qquad \qquad
    - \frac{8505}{16}      \zeta_5     \xil
    - \frac{4635}{32}      \zeta_5     \xil^{2}
    + \frac{405}{16}      \zeta_5     \xil^{3}
    + \frac{315}{64}      \zeta_5     \xil^{4}
    - \frac{70245}{64}      \zeta_5
    + \frac{290371}{64}      \xil
\right. \Break \left. \qquad \qquad
    + \frac{22287}{16}      \xil^{2}
    + \frac{21141}{64}      \xil^{3}
    + \frac{2547}{64}      \xil^{4}
\right] 
{}
{}.\end{eqnarray}

\subsection{Subtracting the Quark Gluon Vertex}

Using again the renormalization condition that at one-loop level is
identical to the original $\MOMt$ scheme we have:
\begin{equation}
\Lambda_q^{\sMOMq}(-\mu^2) 
=
1
{}.
\end{equation}
This subtracts just the structure proportional to $\gamma_\mu$ which is
present at the tree level and corresponds to a longitudinally
polarized gluon. Similar operations as above then lead to
\begin{equation}
\alvp^{\sMOMq}(\mu) = \alvp^{\smsbar}(\mu)
\left[ \frac{\Lambda_q^{\smsbar}(-\mu^2)}
{(1+\Sigma_V^{\smsbar}(-\mu^2)) \sqrt{1+\Pi^{\smsbar}(-\mu^2)} }
\right]^2
\end{equation}
and inserting the $\msbar$ results gives:

\begin{eqnarray}
\label{amomqQCD}
\alvp^{\sMOMq} & = & 
 \alvp\;
+ \;     \alvp^{2}\;
\left[
    + \frac{169}{12} 
    - \frac{10}{9}      n_f
    + \frac{9}{2}      \xil
    + \frac{3}{4}      \xil^{2}
\right] 
+ \;     \alvp^{3}\;
\left[
    + \frac{77035}{144} 
    - \frac{4}{3}      n_f     \zeta_3
    - 5     n_f     \xil
\right. \Break \left. \qquad \qquad
    - \frac{1913}{27}      n_f
    + \frac{100}{81}      n_f^{2}
    + 18     \zeta_3     \xil
    - 54     \zeta_3
    + \frac{1647}{16}      \xil
    + \frac{549}{16}      \xil^{2}
    + \frac{81}{16}      \xil^{3}
\right]\Break
+ \;     \alvp^{4}\;
\left[
    + \frac{42735097}{1728} 
    - \frac{333}{2}      n_f     \zeta_3     \xil
    + \frac{3}{4}      n_f     \zeta_3     \xil^{2}
    + \frac{37579}{108}      n_f     \zeta_3
    + \frac{2320}{9}      n_f     \zeta_5
    - \frac{6759}{16}      n_f     \xil
\right. \Break \left. \qquad \qquad
    - \frac{757}{16}      n_f     \xil^{2}
    - \frac{386759}{81}      n_f
    + \frac{16}{3}      n_f^{2}     \zeta_3     \xil
    + \frac{28}{9}      n_f^{2}     \zeta_3
    + \frac{38}{9}      n_f^{2}     \xil
    + \frac{199903}{972}      n_f^{2}
\right. \Break \left. \qquad \qquad
    - \frac{1000}{729}      n_f^{3}
    + \frac{20067}{16}      \zeta_3     \xil
    + \frac{2511}{32}      \zeta_3     \xil^{2}
    - \frac{225}{8}      \zeta_3     \xil^{3}
    - \frac{117}{32}      \zeta_3     \xil^{4}
    - \frac{35385}{8}      \zeta_3
\right. \Break \left. \qquad \qquad
    - \frac{8325}{16}      \zeta_5     \xil
    - \frac{3195}{32}      \zeta_5     \xil^{2}
    + \frac{255}{16}      \zeta_5     \xil^{3}
    + \frac{315}{64}      \zeta_5     \xil^{4}
    - \frac{66765}{64}      \zeta_5
    + \frac{281059}{64}      \xil
\right. \Break \left. \qquad \qquad
    + \frac{43311}{32}      \xil^{2}
    + \frac{20925}{64}      \xil^{3}
    + \frac{2547}{64}      \xil^{4}
\right] 
{}
{}.\end{eqnarray}

\subsection{Subtracting the Triple Gluon Vertex I}

Subtracting just the Lorentz structure that is present at the tree
level the gluon vertex gives the following renormalization condition
\begin{equation}
\label{form:rencond}
T^{\sMOMg}_1(-\mu^2)
=
1
{}.
\end{equation}  
This subtracts the triple gluon vertex with the zero momentum gluon
and one of the others being polarized longitudinal (in direction of
their momenta) and the last being polarized transversal and parallel
to the plane defined by the gluon momenta (for details see
\cite{phrva:d24:1369}).  A similar derivation as for the ghost gluon vertex
leads then to the following relation of the coupling in the $\MOMg$
scheme to the $\msbar$ coupling:
\begin{equation}
\alvp^{\sMOMg}(\mu) = \alvp^{\smsbar}(\mu) \frac{ \left(
T^{\smsbar}_1(-\mu^2) \right)^2 }{
\left( 1+\Pi^{\smsbar}(-\mu^2) \right)^3 } 
{}.
\end{equation}

This gives the following relations between the coupling constants of
the two schemes:

\begin{eqnarray}
\label{amomgQCD}
\alvp^{\sMOMg} & = & 
 \alvp\;
+ \;     \alvp^{2}\;
\left[
    + \frac{169}{12} 
    - \frac{10}{9}      n_f
    + \frac{9}{2}      \xil
    + \frac{3}{4}      \xil^{2}
\right] 
+ \;     \alvp^{3}\;
\left[
    + \frac{38261}{72} 
    - \frac{4}{3}      n_f     \zeta_3
    - 5     n_f     \xil
\right. \Break \left. \qquad \qquad
    - \frac{7571}{108}      n_f
    + \frac{100}{81}      n_f^{2}
    + \frac{81}{4}      \zeta_3     \xil
    - \frac{207}{4}      \zeta_3
    + \frac{795}{8}      \xil
    + \frac{549}{16}      \xil^{2}
    + \frac{81}{16}      \xil^{3}
\right]\Break
+ \;     \alvp^{4}\;
\left[
    + \frac{84464417}{3456} 
    - \frac{1375}{8}      n_f     \zeta_3     \xil
    + \frac{3}{4}      n_f     \zeta_3     \xil^{2}
    + \frac{72161}{216}      n_f     \zeta_3
    + \frac{2320}{9}      n_f     \zeta_5
\right. \Break \left. \qquad \qquad
    - \frac{39197}{96}      n_f     \xil
    - \frac{187}{4}      n_f     \xil^{2}
    - \frac{6098639}{1296}      n_f
    + \frac{16}{3}      n_f^{2}     \zeta_3     \xil
    + \frac{28}{9}      n_f^{2}     \zeta_3
    + \frac{38}{9}      n_f^{2}     \xil
\right. \Break \left. \qquad \qquad
    + \frac{197149}{972}      n_f^{2}
    - \frac{1000}{729}      n_f^{3}
    + \frac{87729}{64}      \zeta_3     \xil
    + \frac{6261}{64}      \zeta_3     \xil^{2}
    - \frac{1371}{64}      \zeta_3     \xil^{3}
    - \frac{117}{32}      \zeta_3     \xil^{4}
\right. \Break \left. \qquad \qquad
    - \frac{279861}{64}      \zeta_3
    - \frac{26805}{64}      \zeta_5     \xil
    - \frac{4155}{64}      \zeta_5     \xil^{2}
    + \frac{885}{64}      \zeta_5     \xil^{3}
    + \frac{315}{64}      \zeta_5     \xil^{4}
    - \frac{3285}{4}      \zeta_5
\right. \Break \left. \qquad \qquad
    + \frac{267071}{64}      \xil
    + \frac{83511}{64}      \xil^{2}
    + \frac{41445}{128}      \xil^{3}
    + \frac{2547}{64}      \xil^{4}
\right] 
{}
{}.\end{eqnarray}

\subsection{Subtracting the Triple Gluon Vertex II}

The $\MOMgg$ scheme used in \cite{hep-ph/9810437} and
\cite{hep-ph/9903364} is defined by the following renormalization
conditions:
\begin{equation}
T^{\sMOMgg}_1(-\mu^2)-\frac{1}{2}T^{\sMOMgg}_2(-\mu^2)
=
1
{}.
\end{equation}  
This subtracts the triple gluon vertex with the zero momentum gluon
being polarized transversal and parallel to the plane defined by the
gluon momenta. The other two gluon are also polarized transversal but
perpendicular to the vertex plane. Another possibility to understand
this linear combination is to realize that it corresponds to
subtracting the transversal part of the 3-point function, which is the
only one to survive at the \ZP\ (see also \cite{hep-ph/9810437}):
\begin{equation}
G^{(3)abc}_{\mu\nu\rho}(p,-p,0) = (g_{\mu\nu} - \frac{p_\mu
p_\nu}{p^2})p_\rho G^{(3)}(p^2)
{}
\end{equation}
Inserting
eqs.~(\ref{def:vertices},\ref{def:selfenergies},\ref{form:gggdecomposition})
shows that indeed:
\begin{eqnarray}
G^{(3) abc}_{\mu_1\mu_2\mu_3}(p,-p,0) & = & 
(g_{\mu_1\mu_2} - \frac{p_{\mu_1} p_{\mu_2}}{p^2})p_{\mu_3} 
\frac{
\left(T_1(p^2)-\frac{1}{2}T_2(p^2)\right)}{(1+\Pi^{(2)}(p^2))^2
(1+\Pi^{(2)}(0))}
{}.
\end{eqnarray}
From this follows the relation of the coupling in the $\MOMgg$ scheme
to the $\msbar$ coupling:
\begin{equation}
\alvp^{\sMOMgg}(\mu) = \alvp^{\smsbar}(\mu) \frac{ \left(
T^{\smsbar}_1(-\mu^2) - \frac{1}{2} T^{\smsbar}_2(\mu^2) \right)^2 }{
\left( 1+\Pi^{\smsbar}(-\mu^2) \right)^3 } 
{}.
\end{equation}

Inserting $T_1$, $T_2$ and $\Pi$ gives 

\begin{eqnarray}
\label{amomggQCD}
\alvp^{\sMOMgg} & = & 
 \alvp\;
+ \;     \alvp^{2}\;
\left[
    + \frac{70}{3} 
    - \frac{22}{9}      n_f
\right] 
+ \;     \alvp^{3}\;
\left[
    + \frac{516217}{576} 
    - \frac{4}{3}      n_f     \zeta_3
    - \frac{7}{4}      n_f     \xil
    + \frac{1}{2}      n_f     \xil^{2}
\right. \Break \left. \qquad \qquad
    - \frac{8125}{54}      n_f
    + \frac{376}{81}      n_f^{2}
    + \frac{63}{4}      \zeta_3     \xil
    - \frac{153}{4}      \zeta_3
    + \frac{225}{8}      \xil
    + \frac{249}{32}      \xil^{2}
    + \frac{45}{16}      \xil^{3}
    + \frac{9}{64}      \xil^{4}
\right]\Break
+ \;     \alvp^{4}\;
\left[
    + \frac{304676635}{6912} 
    - \frac{1197}{8}      n_f     \zeta_3     \xil
    - \frac{21}{8}      n_f     \zeta_3     \xil^{2}
    + \frac{13339}{27}      n_f     \zeta_3
    - 15     n_f     \zeta_5     \xil
\right. \Break \left. \qquad \qquad
    + \frac{1885}{9}      n_f     \zeta_5
    - \frac{16663}{96}      n_f     \xil
    - \frac{23}{48}      n_f     \xil^{2}
    - \frac{3}{32}      n_f     \xil^{3}
    - \frac{7}{32}      n_f     \xil^{4}
    - \frac{13203725}{1296}      n_f
\right. \Break \left. \qquad \qquad
    + \frac{8}{3}      n_f^{2}     \zeta_3     \xil
    + \frac{40}{9}      n_f^{2}     \zeta_3
    + \frac{44}{9}      n_f^{2}     \xil
    - \frac{8}{9}      n_f^{2}     \xil^{2}
    + \frac{580495}{972}      n_f^{2}
    - \frac{5680}{729}      n_f^{3}
\right. \Break \left. \qquad \qquad
    + \frac{138171}{64}      \zeta_3     \xil
    - \frac{63}{64}      \zeta_3     \xil^{2}
    + \frac{423}{64}      \zeta_3     \xil^{3}
    - \frac{27}{32}      \zeta_3     \xil^{4}
    - \frac{299961}{64}      \zeta_3
    - \frac{47295}{64}      \zeta_5     \xil
\right. \Break \left. \qquad \qquad
    + \frac{3825}{64}      \zeta_5     \xil^{2}
    - \frac{1665}{64}      \zeta_5     \xil^{3}
    - \frac{81825}{64}      \zeta_5
    + \frac{188523}{128}      \xil
    + \frac{118591}{256}      \xil^{2}
    + \frac{5175}{32}      \xil^{3}
\right. \Break \left. \qquad \qquad
    + \frac{5829}{256}      \xil^{4}
    - \frac{27}{64}      \xil^{5}
    - \frac{27}{256}      \xil^{6}
\right]
{}.
\Break
\end{eqnarray}

Following these examples and using the results in the appendix, it
should be possible to construct the relations to relate the coupling
constant in any $\MOMt$ renormalization scheme to the $\msbar$ one.

\section{The $\MOMt$ $\beta$-functions}
\label{sec:betamom}

The first two coefficients of the $\beta$-function in any massless
renormalization scheme are scheme independent. When considering
renormalization schemes in which the coupling constant depends on the
generic covariant gauge parameter, this statement is not true, since
the additional gauge dependence spoils these universality just as a
mass parameter \cite{Tarasov:1990ps}. This means that we need not only
to take into account the gauge parameter when writing down
renormalization group equations, but also that there is a difference
between the gauge parameter defined in the $\msbar$ scheme and in any
of the $\MOMt$ schemes we are considering:
\begin{equation}
\xil^{\smsbar} = \frac{Z_3^{\sMOMt}}{Z_3^{\smsbar}} \xil^{\sMOMt} 
{}.
\end{equation}
In Landau gauge we do not need to consider the additional $\mu$
dependence introduced by the gauge parameter nor the difference
between $\xil^{\smsbar}$ and $\xil^{\sMOMt}$. So in Landau gauge the
$\beta$-function in any $\MOMt$ like scheme can be obtained in the
following simple way once the $\MOMt$ coupling is expressed as a
series in the $\msbar$ coupling
\begin{equation}
\beta^{\sMOMt}(\alvp^{\sMOMt})=\mu^2 \frac{\mathrm{d} \alvp^{\sMOMt}
}{\mathrm{d} \mu^2} = \frac{\partial \alvp^{\sMOMt}}{\partial
\alvp^{\smsbar}} \mu^2 \frac{\mathrm{d} \alvp^{\smsbar}}{\mathrm{d} \mu^2} =
\frac{\partial \alvp^{\sMOMt}}{\partial \alvp^{\smsbar}}
\beta^{\smsbar}(\alvp^{\smsbar})
{}.
\end{equation}
where one has to invert the series $\alvp^{\sMOMt}(\alvp^{\smsbar})$ and
insert this series on the right hand side to express the $\MOMt$
$\beta$-function as a series in $\alvp^{\sMOMt}$.  The four-loop QCD
$\beta$-function in the $\msbar$ scheme was obtained in
\cite{phlta:b400:379} and reads:

\begin{eqnarray}
\label{betamsQCD}
\beta^{\smsbar} & = & 
 \alvp^{2}\;
\left[
    - 11
    + \frac{2}{3}      n_f
\right] 
+ \;     \alvp^{3}\;
\left[
    - 102
    + \frac{38}{3}      n_f
\right] 
+ \;     \alvp^{4}\;
\left[
    - \frac{2857}{2} 
    + \frac{5033}{18}      n_f
    - \frac{325}{54}      n_f^{2}
\right]\Break
+ \;     \alvp^{5}\;
\left[
    - \frac{149753}{6} 
    + \frac{6508}{27}      n_f     \zeta_3
    + \frac{1078361}{162}      n_f
    - \frac{6472}{81}      n_f^{2}     \zeta_3
    - \frac{50065}{162}      n_f^{2}
\right. \Break \left. \qquad \qquad
    - \frac{1093}{729}      n_f^{3}
    - 3564     \zeta_3
\right] 
{}
{},\end{eqnarray}

Combining this result with the relations between the coupling
constants, we can finally obtain the first four coefficients of the
$\beta$-functions in the above introduced four $\MOMt$ schemes, where
we use the definition:
\begin{equation}
\beta(\alvp) = \sum_{i=0}^{3} - \alvp^{(i+2)} \beta_i
\end{equation}
and in Landau gauge $\beta_0$ and $\beta_1$ are independent of the
renormalization scheme. The three- and four-loop contribution in
Landau gauge for the four schemes under consideration are:
\begin{eqnarray}
\label{b3momgQCD}
\beta_3^{\sMOMg} & = & 
 \;
\left[
    + \frac{58491}{16} 
    - \frac{2277}{4}      \zeta_3
\right] 
+ \;     n_f\;
\left[
    - \frac{15283}{24} 
    + \frac{119}{6}      \zeta_3
\right] 
+ \;     n_f^{2}\;
\left[
    + \frac{481}{27} 
    + \frac{8}{9}      \zeta_3
\right] 
{}
{},
\end{eqnarray}
\begin{eqnarray}
\label{b4momgQCD}
\beta_4^{\sMOMg} & = & 
 \;
\left[
    + \frac{10982273}{64} 
    - \frac{1425171}{32}      \zeta_3
    - \frac{36135}{2}      \zeta_5
\right] 
+ \;     n_f\;
\left[
    - \frac{3830167}{96} 
    + \frac{1075423}{144}      \zeta_3
    + \frac{60895}{9}      \zeta_5
\right]\Break
+ \;     n_f^{2}\;
\left[
    + \frac{724445}{324} 
    - \frac{12959}{54}      \zeta_3
    - \frac{9280}{27}      \zeta_5
\right] 
+ \;     n_f^{3}\;
\left[
    - \frac{788}{27} 
    + \frac{16}{9}      \zeta_3
\right] 
{}
{},
\end{eqnarray}
\begin{eqnarray}
\label{b3momhQCD}
\beta_3^{\sMOMh} & = & 
 \;
\left[
    + \frac{28965}{8} 
    - \frac{3861}{8}      \zeta_3
\right] 
+ \;     n_f\;
\left[
    - \frac{7715}{12} 
    + \frac{175}{12}      \zeta_3
\right] 
+ \;     n_f^{2}\;
\left[
    + \frac{989}{54} 
    + \frac{8}{9}      \zeta_3
\right] 
{},
\end{eqnarray}
\begin{eqnarray}
\label{b4momhQCD}
\beta_4^{\sMOMh} & = & 
 \;
\left[
    + \frac{1380469}{8} 
    - \frac{625317}{16}      \zeta_3
    - \frac{772695}{32}      \zeta_5
\right] 
+ \;     n_f\;
\left[
    - \frac{970819}{24} 
    + \frac{516881}{72}      \zeta_3
    + \frac{1027375}{144}      \zeta_5
\right]\Break
+ \;     n_f^{2}\;
\left[
    + \frac{736541}{324} 
    - \frac{6547}{27}      \zeta_3
    - \frac{9280}{27}      \zeta_5
\right] 
+ \;     n_f^{3}\;
\left[
    - \frac{800}{27} 
    + \frac{16}{9}      \zeta_3
\right] 
{},
\end{eqnarray}
\begin{eqnarray}
\label{b3momqQCD}
\beta_3^{\sMOMq} & = & 
 \;
\left[
    + \frac{29559}{8} 
    - 594     \zeta_3
\right] 
+ \;     n_f\;
\left[
    - \frac{7769}{12} 
    + \frac{64}{3}      \zeta_3
\right] 
+ \;     n_f^{2}\;
\left[
    + \frac{989}{54} 
    + \frac{8}{9}      \zeta_3
\right] 
{},
\end{eqnarray}
\begin{eqnarray}
\label{b4momqQCD}
\beta_4^{\sMOMq} & = & 
 \;
\left[
    + \frac{2795027}{16} 
    - \frac{174207}{4}      \zeta_3
    - \frac{734415}{32}      \zeta_5
\right] 
+ \;     n_f\;
\left[
    - \frac{487751}{12} 
    + \frac{67939}{9}      \zeta_3
    + \frac{1016935}{144}      \zeta_5
\right]\Break
+ \;     n_f^{2}\;
\left[
    + \frac{737837}{324} 
    - \frac{6709}{27}      \zeta_3
    - \frac{9280}{27}      \zeta_5
\right] 
+ \;     n_f^{3}\;
\left[
    - \frac{800}{27} 
    + \frac{16}{9}      \zeta_3
\right] 
{},
\end{eqnarray}
\begin{eqnarray}
\label{b3momggQCD}
\beta_3^{\sMOMgg} & = & 
 \;
\left[
    + \frac{186747}{64} 
    - \frac{1683}{4}      \zeta_3
\right] 
+ \;     n_f\;
\left[
    - \frac{35473}{96} 
    + \frac{65}{6}      \zeta_3
\right] 
+ \;     n_f^{2}\;
\left[
    - \frac{829}{54} 
    + \frac{8}{9}      \zeta_3
\right] 
+ \;     n_f^{3}\;
\left[
    + \frac{8}{9} 
\right]\Break
{},
\end{eqnarray}
\begin{eqnarray}
\label{b4momggQCD}
\beta_4^{\sMOMgg} & = & 
 \;
\left[
    + \frac{20783939}{128} 
    - \frac{1300563}{32}      \zeta_3
    - \frac{900075}{32}      \zeta_5
\right]\Break
+ \;     n_f\;
\left[
    - \frac{2410799}{64} 
    + \frac{1323259}{144}      \zeta_3
    + \frac{908995}{144}      \zeta_5
\right]\Break
+ \;     n_f^{2}\;
\left[
    + \frac{1464379}{648} 
    - \frac{12058}{27}      \zeta_3
    - \frac{7540}{27}      \zeta_5
\right]\Break
+ \;     n_f^{3}\;
\left[
    - \frac{3164}{27} 
    + \frac{64}{9}      \zeta_3
\right] 
+ \;     n_f^{4}\;
\left[
    + \frac{320}{81} 
\right] 
{}.
\end{eqnarray}

\section{Discussion}
\label{discussion}

In numerical form the coupling constant relations and the
corresponding $\beta$-functions for these four schemes in Landau gauge
read:

\begin{eqnarray}
\label{amomhnum}
\alvp^{\sMOMh} & = & 
 \alvp\;
+ \;     \alvp^{2}\;
\left[
+  14.0833\,
-  1.11111\,     n_f
\right] 
+ \;     \alvp^{3}\;
\left[
+  475.475\,
-  72.4546\,     n_f
+  1.23457\,     n_f^{2}
\right]\Break
+ \;     \alvp^{4}\;
\left[
+  18652.4\,
-  4109.72\,     n_f
+  209.401\,     n_f^{2}
-  1.37174\,     n_f^{3}
\right] 
{},\end{eqnarray}
\begin{eqnarray}
\label{amomqnum}
\alvp^{\sMOMq} & = & 
 \alvp\;
+ \;     \alvp^{2}\;
\left[
+  14.0833\,
-  1.11111\,     n_f
\right] 
+ \;     \alvp^{3}\;
\left[
+  470.054\,
-  72.4546\,     n_f
+  1.23457\,     n_f^{2}
\right]\Break
+ \;     \alvp^{4}\;
\left[
+  18332.4\,
-  4089.25\,     n_f
+  209.401\,     n_f^{2}
-  1.37174\,     n_f^{3}
\right] 
{},\end{eqnarray}
\begin{eqnarray}
\label{amomgnum}
\alvp^{\sMOMg} & = & 
 \alvp\;
+ \;     \alvp^{2}\;
\left[
+  14.0833\,
-  1.11111\,     n_f
\right] 
+ \;     \alvp^{3}\;
\left[
+  469.196\,
-  71.7046\,     n_f
+  1.23457\,     n_f^{2}
\right]\Break
+ \;     \alvp^{4}\;
\left[
+  18332\,
-  4036.86\,     n_f
+  206.568\,     n_f^{2}
-  1.37174\,     n_f^{3}
\right] 
{},\end{eqnarray}
\begin{eqnarray}
\label{amomggnum}
\alvp^{\sMOMgg} & = & 
 \alvp\;
+ \;     \alvp^{2}\;
\left[
+  23.3333\,
-  2.44444\,     n_f
\right]\Break
+ \;     \alvp^{3}\;
\left[
+  850.231\,
-  152.066\,     n_f
+  4.64198\,     n_f^{2}
\right]\Break
+ \;     \alvp^{4}\;
\left[
+  37119.7\,
-  9377.02\,     n_f
+  602.56\,     n_f^{2}
-  7.7915\,     n_f^{3}
\right] 
{},
\end{eqnarray}
\begin{eqnarray}
\label{bmomhnum}
\beta^{\sMOMh} & = & 
 \left(\alvp^{\sMOMh}\right)^{2}\;
\left[
-  11\,
+ 0.666667\,     n_f
\right] 
+ \;     \left(\alvp^{\sMOMh}\right)^{3}\;
\left[
-  102\,
+  12.6667\,     n_f
\right]\Break
+ \;     \left(\alvp^{\sMOMh}\right)^{4}\;
\left[
-  3040.48\,
+  625.387\,     n_f
-  19.3833\,     n_f^{2}
\right]\Break
+ \;     \left(\alvp^{\sMOMh}\right)^{5}\;
\left[
-   100541\,
+  24423.3\,     n_f
-  1625.4\,     n_f^{2}
+  27.4926\,     n_f^{3}
\right] 
{},
\end{eqnarray}
\begin{eqnarray}
\label{bmomgnum}
\beta^{\sMOMg} & = & 
 \left(\alvp^{\sMOMg}\right)^{2}\;
\left[
-  11\,
+ 0.666667\,     n_f
\right] 
+ \;     \left(\alvp^{\sMOMg}\right)^{3}\;
\left[
-  102\,
+  12.6667\,     n_f
\right]\Break
+ \;     \left(\alvp^{\sMOMg}\right)^{4}\;
\left[
-  2971.42\,
+  612.951\,     n_f
-  18.8833\,     n_f^{2}
\right]\Break
+ \;     \left(\alvp^{\sMOMg}\right)^{5}\;
\left[
-  99327.8\,
+  23904.4\,     n_f
-  1591.07\,     n_f^{2}
+  27.0482\,     n_f^{3}
\right] 
{},
\end{eqnarray}
\begin{eqnarray}
\label{bmomqnum}
\beta^{\sMOMq} & = & 
 \left(\alvp^{\sMOMq}\right)^{2}\;
\left[
-  11\,
+ 0.666667\,     n_f
\right] 
+ \;     \left(\alvp^{\sMOMq}\right)^{3}\;
\left[
-  102\,
+  12.6667\,     n_f
\right]\Break
+ \;     \left(\alvp^{\sMOMq}\right)^{4}\;
\left[
-  2980.85\,
+  621.773\,     n_f
-  19.3833\,     n_f^{2}
\right]\Break
+ \;     \left(\alvp^{\sMOMq}\right)^{5}\;
\left[
-  98539.5\,
+  24249\,     n_f
-  1622.19\,     n_f^{2}
+  27.4926\,     n_f^{3}
\right] 
{},
\end{eqnarray}
\begin{eqnarray}
\label{bmomggnum}
\beta^{\sMOMgg} & = & 
 \left(\alvp^{\sMOMgg}\right)^{2}\;
\left[
-  11\,
+ 0.666667\,     n_f
\right] 
+ \;     \left(\alvp^{\sMOMgg}\right)^{3}\;
\left[
-  102\,
+  12.6667\,     n_f
\right]\Break
+ \;     \left(\alvp^{\sMOMgg}\right)^{4}\;
\left[
-  2412.16\,
+  356.488\,     n_f
+  14.2834\,     n_f^{2}
- 0.888889\,     n_f^{3}
\right]\Break
+ \;     \left(\alvp^{\sMOMgg}\right)^{5}\;
\left[
-  84353.8\,
+  20077.1\,     n_f
-  1433.44\,     n_f^{2}
+  108.637\,     n_f^{3}
-  3.95062\,     n_f^{4}
\right]
{}.
\Break
\end{eqnarray}

The corresponding numbers for the $\msbar$ $\beta$-function are (as an
expansion in $\alvp=\alvp^{\smsbar}$):

\begin{eqnarray}
\label{bmsnum}
\beta^{\smsbar} & = & 
 \alvp^{2}\;
\left[
-  11\,
+ 0.666667\,     n_f
\right] 
+ \;     \alvp^{3}\;
\left[
-  102\,
+  12.6667\,     n_f
\right]\Break
+ \;     \alvp^{4}\;
\left[
-  1428.5\,
+  279.611\,     n_f
-  6.01852\,     n_f^{2}
\right]\Break
+ \;     \alvp^{5}\;
\left[
-  29243\,
+  6946.29\,     n_f
-  405.089\,     n_f^{2}
-  1.49931\,     n_f^{3}
\right] 
{}
{}.\end{eqnarray}

It has been argued~\cite{phrva:d20:1420} that momentum subtraction
schemes will lead to couplings that are only slightly
different. Comparing the $\MOMgg$ scheme with the other three schemes
it is easy to see that this of course depends a lot on the linear
combination used in the renormalization condition. On the other hand
it is clearly seen that the couplings as well as the $\beta$-functions
of the three schemes that are identical at one-loop order will also
give very close-by values for the coupling constant at higher
orders. Also the $\MOMgg$ scheme is still much
closer to the other $\MOMt$ schemes than to the $\msbar$ one.

Inserting the values $1,\dots,6$ for $n_f$ shows that the coefficients
for all the values of $n_f$ are smaller for the $\msbar$ scheme than
in any of the $\MOMt$ schemes.  The difference between the $\msbar$
scheme and the $\MOMt$ schemes is only small for $n_f=6$ and gets
quite large for the small values of $n_f$.

Recently the momentum dependence (running) of the three-gluon
asymmetrical vertex corresponding to the  combination $T_{gg} =
T_1(p^2) - \frac{1}{2} T_2$ was computed within the lattice approach
in \cite{Boucaud:2000ey}. The resulting (nonperturbative!) behaviour
of 
\[
\alpha_s(p^2) = 4 \pi\alvp^{\sMOMgg}(p^2) 
\] 
in the flavorless QCD has been found 
to be best described by an
ansatz\footnote{To be consistent to \cite{Boucaud:2000ey} we use the
Euclidean metrics below.}
\begin{equation}
\alpha_s(p^2) = \alpha_s^{Pert}(p^2) + \frac{c}{p^2}
{},
\end{equation}
with
\begin{equation}
\Lambda_{\msbar} =  237 \pm 3 \mbox{MeV} ,
\ \ \ 
 c = 0.63 \pm 0.03  \mbox{GeV}^2 
\label{}
{}.
\end{equation}
Here $\Lambda_{\msbar} = {\mathrm exp}(-70/66) \Lambda_{\sMOMgg}$.
The above results have been obtained by working at three-loop level
and employing the $\MOMgg$ scheme. The authors of
\cite{Boucaud:2000ey} have also investigated the dependence of the
result on the (then unknown) four-loop contribution to
$\beta^{\sMOMgg}$. Their findings are summarized in Table 1, which we
have copied from \cite{Boucaud:2000ey} with adding one more row
obtained with the help of the linear extrapolation and corresponding to
the true value of the parameter
\[
\frac{b_3}{b_2} = \frac{\beta^{\sMOMgg}_3/(4\pi)}{\beta^{\sMOMgg}_2}
= 2.78284\dots  \  \ \ \  \ \mbox{(for $n_f = 0 $)} 
{}. 
\]   
Table 1 clearly demonstrates that taking into account the four-loop
term in the $\beta^{\sMOMgg}$-function leads to a significant decrease
(around 30\%) of the value of the non-perturbative $1/p^2$ correction.
\begin{table}
\centering
\begin{tabular}{|c|c|c|c|c|c|c|c|c|c|c|}
\hline
\hline
& \multicolumn{4}{c|}{4 loops} & \multicolumn{6}{c|}{4 loops + power} \\
\cline{2-11}
$b_3/b_2$ & \multicolumn{2}{c|}{whole} & \multicolumn{2}{c|}{$>3$}& 
\multicolumn{3}{c|}{whole} & \multicolumn{3}{c|}{$>3$} \\ 
\cline{2-11}
& $\chi^2_{dof}$ & $\Lams$ & $\chi^2_{dof}$ & $\Lams$ 
& $\chi^2_{dof}$ & $\Lams$ & $c$ & $\chi^2_{dof}$ & $\Lams$ & $c$ \\
\hline
0 & 7.8 & 299 & 8.3 & 294 & 1.6 & 237 & 0.63 & 1.6 & 235 & 0.67 \\
1 & 7.1 & 284 & 6.0 & 287 & 1.8 & 238 & 0.56 & 1.6 & 235 & 0.61 \\
2 & 28  & 259 & 4.1 & 279 & 2.7 & 229 & 0.57 & 1.6 & 236 & 0.53 \\
\hline
\hline
2.78 & 87.5 & 234. & 3.2 & 272. & 4.3 & 218. & 0.63 & 1.6 & 238. & 0.47 \\
\hline
\hline
3 & 104 & 227 & 3.0 & 270 & 4.7 & 215 & 0.65 & 1.6 & 238 & 0.45 \\
4 & 145 & 215 & 2.9 & 261 & 7.3 & 202 & 0.75 & 1.8 & 237 & 0.37 \\
5 & 157 & 209 & 4.4 & 252 & 10.2 & 190 & 0.84 & 2.4 & 231 & 0.37 \\
\hline
\hline
\end{tabular}
\caption{\small {\it A collection of fitted parameters obtained by 
imposing different values for the ratio $b_3/b_2$ as defined in the text. 
``Whole'' refers to the whole energy window (2 GeV $< \mu < $10 GeV) 
and ``$>3$'' to a momentum range above $3$ GeV. 
}}
\label{Tab1}
\end{table}             

\section{Conclusions}

In this paper we have analytically computed the full set of the
three-loops propagators and fundamental three-linear vertexes with one
of three external momenta set to zero in the massless QCD.  The
results have been used to find the NNNLO conversion factors
transforming the $\overline{\mathrm{MS}}$ coupling constant to the
ones defined corresponding to a set of regularization independent
renormalization schemes based on momentum subtractions
($\MOMt$-schemes).  Then we have used the conversion factors to
evaluate the four-loop $\beta$ functions in these schemes.

The newly computed corrections to the coupling constant running in
$\MOMt$-schemes prove to be  numerically significant. They should be taken into
account when confronting the running obtained with the help of lattice
simulations with  the pQCD predictions. 
\vspace{1.0cm}

{\bf Acknowledgments} 
\vspace{0.5cm}

K.~G.~Ch.  is grateful  A.~I.~Davydychev and J.H. K\"uhn for
useful discussions. 
A.~R. wants to thank Thorsten Seidensticker, Timo van Ritbergen and
Dominik St\"ockinger for valuable discussions. Thorsten Seidensticker
also provided some preliminary results of work in progress for
crosschecks.  This work was supported by the DFG under Contract Ku
502/8-1 ({\it DFG-Forschergruppe ``Quantenfeldtheorie, Computeralgebra
und Monte-Carlo-Simu\-lationen'' }) and the {\it Graduiertenkolleg
``Elementarteilchenphysik an Beschleunigern'' at the Universit\"at
Karlsruhe}.


\newpage

\begin{appendix}

\numberwithin{equation}{section}

\section{Technical Remarks}
\label{sec:setup}

To evaluate the self-energies and vertex function at three loop order
we made heavy use of computer programs. The program {\tt
QGRAF}~\cite{prog:qgraf} was used to generate the diagrams. For the
self-energies and the vertex functions at the \ZP\ we arrive at scalar
integrals of the massless propagator type after applying some 
projectors. An algorithm which allows
one to analytically evaluate divergent as well as finite parts of such 
integrals was  elaborated in \cite{nupha:b192:159}. 
It has been  implemented in an
efficient way in the {\tt MINCER}~\cite{prog:mincer} package written
for the symbolic manipulation program {\tt FORM}~\cite{prog:form}. The
huge number of diagrams in the three-loop calculations requires  a
complete automation of the whole procedure which should include also the
calculation of the color factors. This has been implemented and is
described in more detail along with other similar installations in
\cite{hep-ph/9812357}. This setup also gives the user tools to perform
series expansions in small parameters, a feature that was used to
calculate the naive first order expansion in a small external momentum
for the ghost vertex which is necessary to check the WST identity as
described in Section~\ref{sec:WSTI}. We had to evaluate the following
diagrams:
\begin{center}
\begin{tabular}{|c|c|c|c|c|}
\hline
Number of diagrams for & 1 loops & 2 loops & 3 loops & approximate
runtime per function\\
\hline
\hline
$\tilde \Pi$,$\Sigma_V$ & 1 & 7 & 106 & 6 hours \\
\hline
$\Pi$ & 4 & 27 & 494 & 12 hours \\
\hline
$2\,\times\,\tilde \Gamma_i+ 4\,\times\, a^{(\prime)}_{i}$ 
& 2 & 40 & 1022 & 12 hours  \\
\hline
$2\,\times\,(\Lambda_i,\Lambda_i^T)$ & 2 & 40 & 1022 & 2 days \\
\hline
$T_i$ & 10 & 189 & 5526 & 4 weeks \\
\hline
\end{tabular}
\end{center}
All calculations were done on workstations using the Alpha 21164
processors running at 600 MHz.  The approximate runtimes are only
rough estimates and show that the calculation of the triple gluon
vertex was by far the most demanding part.  This is not only because
of the number of diagrams but also because the complicated vertex and
propagator structures generate a huge number of intermediate terms.

Wherever possible, we have cross-checked the following expressions
with the known two- and three-loop results, the gluon, ghost and quark
field anomalous dimensions can also be found in \cite{phlta:B93:429}
and \cite{hep-ph/9302208}. In most cases we found complete
agreement. We only find different values than \cite{phrva:d24:1369}
for the two-loop contributions to $\Lambda_g$ and $\Lambda_g^T$ (their
$\Gamma_3$ and $\Gamma_4$). For the two-loop triple gluon vertex we find
complete agreement with \cite{hep-ph/9801380}, which contains a list
of miss-prints in earlier publications, among them the gauge
dependence of the one-loop $T_2$ in \cite{phrva:d24:1369}.

The latex code for all results in this paper has been created
automatically. The code has then been retransformed to {\sc form}
input files and cross-checked against the original expressions. We hope
that by doing these checks and avoiding any hand editing of the
formulas we have circumvent the common problem of miss-prints in
otherwise correct results. The price is of course that the layout of
some of the formulas is not as fancy as it could be. Also note that
these expressions will be made available in the World-Wide-Web.

\section{Conventions}

    \subsection{Color Factors}

The following results are given for generic color factors and are
valid for any semi-simple compact Lie group, where $T^a_{ij}$ are the
generators of the fundamental representation and $f^{abc}$ are the
structure constants of the Lie algebra.
\begin{center}
\begin{tabular}{|c||c|c|c|c|}
\hline
 & Definition & $\SU(N)$ & QCD & QED \\
\hline
\hline
\raisebox{-1.5ex}{\rule{0ex}{4ex}}
$C_A$ & $f^{acd}f^{bcd} = C_A \delta^{ab}$ & $N$ & $3$ & 0\\
\hline
\raisebox{-1.5ex}{\rule{0ex}{4ex}}
$C_F$ & $[T^a T^a]_{ij} = C_F \delta_{ij}$ & $\frac{N^2-1}{2 N}$ &
 $\frac{4}{3}$ & $1$ \\
\hline
\raisebox{-1.5ex}{\rule{0ex}{4ex}}
$T$   & $\mathrm{Tr}(T^a T^b) = T \delta^{ab}$ & $\frac{1}{2}$ &
 $\frac{1}{2}$ & $1$ \\
\hline
\end{tabular}
\end{center}

    \subsection{Momentum Dependence}

We give the $\msbar$ renormalized result for all self-energies and
vertex functions at the point $p^2=-\mu^2$. The correct $p^2$
dependence for any of the self-energies and vertex functions $\Gamma$
can be restored from the renormalization group equations, which are
with our conventions:
\begin{equation}
\mu^2 \frac{\mathrm{d}}{\mathrm{d} \mu^2}
\Gamma(\alvp,q,\mu,\xil) = 
\left( \mu^2 \frac{\partial}{\partial\mu^2}
+ \beta\frac{\partial}{\partial \alvp} 
+ \gamma_3 \xil \frac{\partial}{\partial \xil }\right) 
\Gamma(\alvp,q,\mu,\xil) = 
\gamma_\Gamma  \Gamma(\alvp,q,\mu,\xil) 
\label{RGgen}
\end{equation}
For massless QCD one cam write 
\begin{equation}
\Gamma = \Gamma_0 + \sum_{i  >  0} \Gamma_i  \log^i(\frac{\mu^2}{-q^2})
{},
\end{equation}
and as a direct consequence of (\ref{RGgen}) one gets (for $n >  0)  $ 
\begin{equation}
\Gamma_n = 
\left( - \beta \frac{\partial}{\partial \alvp}
- \gamma_3 \xil \frac{\partial}{\partial \xil}
+ \gamma_\Gamma \right)
 \frac{1}{n}\Gamma_{n-1} 
\end{equation}
The equation can be directly  used to reconstruct the $q$-dependence
of the scalar functions $1+\Pi$, $1+\tilde\Pi$ and $1+\Sigma_V$. For
the vertex functions one needs to multiply the functions $T_i$,
$\tilde \Gamma_i$ and $\Lambda_i$ by $g = 4 \pi \sqrt{\alvp}$. The
anomalous dimensions of any $\Gamma$ is:
\begin{equation}
\gamma_\Gamma = - \left( \frac{n_g}{2}\gamma_3 
+ \frac{n_h}{2}\tilde \gamma_3 + \frac{n_q}{2}\gamma_2 \right)
\end{equation}
where $n_g$,$n_h$ and $n_q$ are the number of external gluon, ghost
and quark fields, respectively, of the diagrams contributing to
$\Gamma$. The QCD $\beta$-function and the field anomalous dimensions
are given in part \ref{app:anomdims} of this appendix. Note that in
our definitions for the anomalous dimensions a relative sign and a
factor 2 is different from many other publications.

\section{Propagators and Vertices in massless QCD}
\label{app:masslessQCD}

   \subsection{The Gluon Self-Energy}
\label{app:gg}

\begin{eqnarray}
\label{PirenMSb}
\Pi^{\smsbar} & = & 
 \alvp C_A\;
\left[
    - \frac{97}{36} 
    - \frac{1}{2}      \xil
    - \frac{1}{4}      \xil^{2}
\right] 
+ \;     \alvp     T     n_f\;
\left[
    + \frac{20}{9} 
\right]\Break
+ \;     \alvp^{2}     C_A     T     n_f\;
\left[
    + \frac{59}{4} 
    + 8     \zeta_3
    - \frac{10}{9}      \xil
    - \frac{10}{9}      \xil^{2}
\right] 
+ \;     \alvp^{2}     C_F     T     n_f\;
\left[
    + \frac{55}{3} 
    - 16     \zeta_3
\right]\Break
+ \;     \alvp^{2}     C_A^{2}\;
\left[
    - \frac{2381}{96} 
    - 2     \zeta_3     \xil
    + 3     \zeta_3
    + \frac{463}{288}      \xil
    + \frac{95}{144}      \xil^{2}
    - \frac{1}{16}      \xil^{3}
    + \frac{1}{16}      \xil^{4}
\right]\Break
+ \;     \alvp^{3}     C_A^{3}\;
\left[
    - \frac{10221367}{31104} 
    - \frac{12071}{288}      \zeta_3     \xil
    - \frac{161}{96}      \zeta_3     \xil^{2}
    + \frac{149}{96}      \zeta_3     \xil^{3}
    + \frac{13}{96}      \zeta_3     \xil^{4}
    + \frac{1549}{24}      \zeta_3
    + \frac{3}{8}      \zeta_4     \xil
\right. \Break \left. \qquad \qquad
    + \frac{3}{32}      \zeta_4     \xil^{2}
    + \frac{9}{32}      \zeta_4
    + \frac{115}{8}      \zeta_5     \xil
    + \frac{385}{96}      \zeta_5     \xil^{2}
    - \frac{5}{24}      \zeta_5     \xil^{3}
    - \frac{35}{192}      \zeta_5     \xil^{4}
    + \frac{7025}{192}      \zeta_5
\right. \Break \left. \qquad \qquad
    + \frac{13141}{1152}      \xil
    + \frac{30835}{10368}      \xil^{2}
    - \frac{2813}{1152}      \xil^{3}
    - \frac{29}{48}      \xil^{4}
    + \frac{1}{16}      \xil^{5}
    - \frac{1}{64}      \xil^{6}
\right]\Break
+ \;     \alvp^{3}     C_A^{2}     T     n_f\;
\left[
    + \frac{1154561}{3888} 
    + \frac{202}{9}      \zeta_3     \xil
    - \frac{25}{6}      \zeta_3     \xil^{2}
    + \frac{871}{18}      \zeta_3
    - 9     \zeta_4
    - \frac{160}{3}      \zeta_5
    - \frac{241}{24}      \xil
\right. \Break \left. \qquad \qquad
    - \frac{6137}{1296}      \xil^{2}
    - \frac{5}{18}      \xil^{3}
    + \frac{5}{12}      \xil^{4}
\right]\Break
+ \;     \alvp^{3}     C_A     T^{2}     n_f^{2}\;
\left[
    - \frac{10499}{243} 
    - \frac{64}{9}      \zeta_3     \xil
    - \frac{256}{9}      \zeta_3
    + \frac{16}{9}      \xil
    - \frac{100}{81}      \xil^{2}
\right]\Break
+ \;     \alvp^{3}     C_A     C_F     T     n_f\;
\left[
    + \frac{96809}{324} 
    + 8     \zeta_3     \xil
    + 8     \zeta_3     \xil^{2}
    - \frac{1492}{9}      \zeta_3
    + 12     \zeta_4
    - 80     \zeta_5
    - \frac{55}{6}      \xil
    - \frac{55}{6}      \xil^{2}
\right]\Break
+ \;     \alvp^{3}     C_F     T^{2}     n_f^{2}\;
\left[
    - \frac{7402}{81} 
    + \frac{608}{9}      \zeta_3
\right] 
+ \;     \alvp^{3}     C_F^{2}     T     n_f\;
\left[
    - \frac{286}{9} 
    - \frac{296}{3}      \zeta_3
    + 160     \zeta_5
\right] 
{}
{}.\end{eqnarray}

   \subsection{The Ghost Self-Energy}
\label{app:cc}

\begin{eqnarray}
\label{PitildeMSb}
\tilde\Pi^{\smsbar} & = & 
 \alvp C_A\;
\left[
    - 1
\right] 
+ \;     \alvp^{2}     C_A     T     n_f\;
\left[
    + \frac{95}{24} 
\right]\Break
+ \;     \alvp^{2}     C_A^{2}\;
\left[
    - \frac{1751}{192} 
    - \frac{3}{8}      \zeta_3     \xil
    + \frac{3}{16}      \zeta_3     \xil^{2}
    + \frac{15}{16}      \zeta_3
    + \frac{7}{64}      \xil
    - \frac{3}{8}      \xil^{2}
\right]\Break
+ \;     \alvp^{3}     C_A^{3}\;
\left[
    - \frac{466373}{3888} 
    - \frac{1429}{192}      \zeta_3     \xil
    + \frac{93}{64}      \zeta_3     \xil^{2}
    + \frac{23}{64}      \zeta_3     \xil^{3}
    + \frac{12403}{576}      \zeta_3
    - \frac{3}{16}      \zeta_4     \xil
    - \frac{3}{64}      \zeta_4     \xil^{2}
\right. \Break \left. \qquad \qquad
    - \frac{9}{64}      \zeta_4
    + \frac{65}{32}      \zeta_5     \xil
    - \frac{35}{32}      \zeta_5     \xil^{2}
    + \frac{5}{32}      \zeta_5     \xil^{3}
    + \frac{65}{32}      \zeta_5
    - \frac{61}{576}      \xil
    - \frac{327}{128}      \xil^{2}
    - \frac{361}{384}      \xil^{3}
\right]\Break
+ \;     \alvp^{3}     C_A^{2}     T     n_f\;
\left[
    + \frac{150247}{1944} 
    + \frac{13}{6}      \zeta_3     \xil
    + \frac{29}{9}      \zeta_3
    + \frac{9}{2}      \zeta_4
    - \frac{199}{144}      \xil
\right]\Break
+ \;     \alvp^{3}     C_A     T^{2}     n_f^{2}\;
\left[
    - \frac{5161}{486} 
    - \frac{8}{9}      \zeta_3
\right] 
+ \;     \alvp^{3}     C_A     C_F     T     n_f\;
\left[
    + \frac{899}{24} 
    - 22     \zeta_3
    - 6     \zeta_4
\right] 
{}
{}.\end{eqnarray}

   \subsection{The Quark Self-Energy}
\label{app:qq}

\begin{eqnarray}
\label{SigmaVrenMSb}
\Sigma_V^{\smsbar} & = & 
 \alvp C_F\;
\left[
    + \xil
\right] 
+ \;     \alvp^{2}     C_A     C_F\;
\left[
    + \frac{41}{4} 
    - 3     \zeta_3     \xil
    - 3     \zeta_3
    + \frac{13}{2}      \xil
    + \frac{9}{8}      \xil^{2}
\right]\Break
+ \;     \alvp^{2}     C_F     T     n_f\;
\left[
    - \frac{7}{2} 
\right] 
+ \;     \alvp^{2}     C_F^{2}\;
\left[
    - \frac{5}{8} 
\right]\Break
+ \;     \alvp^{3}     C_A^{2}     C_F\;
\left[
    + \frac{159257}{648} 
    - 35     \zeta_3     \xil
    - \frac{39}{8}      \zeta_3     \xil^{2}
    - \frac{1}{3}      \zeta_3     \xil^{3}
    - \frac{3139}{24}      \zeta_3
    + \frac{3}{8}      \zeta_4     \xil
    + \frac{3}{16}      \zeta_4     \xil^{2}
\right. \Break \left. \qquad \qquad
    - \frac{69}{16}      \zeta_4
    + \frac{5}{2}      \zeta_5     \xil
    + \frac{5}{4}      \zeta_5     \xil^{2}
    + \frac{165}{4}      \zeta_5
    + \frac{39799}{576}      \xil
    + \frac{787}{64}      \xil^{2}
    + \frac{55}{24}      \xil^{3}
\right]\Break
+ \;     \alvp^{3}     C_A     C_F^{2}\;
\left[
    - \frac{997}{24} 
    - 17     \zeta_3     \xil
    + \zeta_3     \xil^{3}
    + 44     \zeta_3
    + 6     \zeta_4
    + 20     \zeta_5     \xil
    - 20     \zeta_5
    + 4     \xil
    + \frac{3}{2}      \xil^{2}
    - \frac{1}{8}      \xil^{3}
\right]\Break
+ \;     \alvp^{3}     C_A     C_F     T     n_f\;
\left[
    - \frac{11887}{81} 
    + 8     \zeta_3     \xil
    + \frac{52}{3}      \zeta_3
    - \frac{1723}{72}      \xil
\right] 
+ \;     \alvp^{3}     C_F     T^{2}     n_f^{2}\;
\left[
    + \frac{1570}{81} 
\right]\Break
+ \;     \alvp^{3}     C_F^{2}     T     n_f\;
\left[
    - \frac{79}{6} 
    + 16     \zeta_3
    - \frac{3}{2}      \xil
\right] 
+ \;     \alvp^{3}     C_F^{3}\;
\left[
    - \frac{73}{12} 
    - \frac{2}{3}      \zeta_3     \xil^{3}
    + \frac{7}{8}      \xil
\right] 
{}
{}.\end{eqnarray}

   \subsection{The Triple Gluon Vertex}
\label{app:ggg}

\begin{eqnarray}
\label{T1renMSb}
T_1^{\smsbar} & = & 
1\;
{}
+ \; \alvp C_A\;
\left[
    - \frac{61}{36} 
    - \frac{1}{4}      \xil^{2}
\right] 
+ \;     \alvp     T     n_f\;
\left[
    + \frac{20}{9} 
\right]\Break
+ \;     \alvp^{2}     C_A^{2}\;
\left[
    - \frac{9907}{576} 
    - \frac{15}{8}      \zeta_3     \xil
    + \frac{13}{8}      \zeta_3
    + \frac{153}{64}      \xil
    + \frac{35}{36}      \xil^{2}
    - \frac{3}{16}      \xil^{3}
    + \frac{1}{16}      \xil^{4}
\right]\Break
+ \;     \alvp^{2}     C_A     T     n_f\;
\left[
    + \frac{955}{72} 
    + 8     \zeta_3
    - \frac{10}{9}      \xil^{2}
\right] 
+ \;     \alvp^{2}     C_F     T     n_f\;
\left[
    + \frac{55}{3} 
    - 16     \zeta_3
\right]\Break
+ \;     \alvp^{3}     C_A^{3}\;
\left[
    - \frac{155555}{648} 
    - \frac{48047}{1152}      \zeta_3     \xil
    - \frac{1481}{384}      \zeta_3     \xil^{2}
    + \frac{145}{128}      \zeta_3     \xil^{3}
    + \frac{13}{96}      \zeta_3     \xil^{4}
    + \frac{68537}{1152}      \zeta_3
\right. \Break \left. \qquad \qquad
    + \frac{9}{16}      \zeta_4     \xil
    + \frac{9}{64}      \zeta_4     \xil^{2}
    + \frac{27}{64}      \zeta_4
    + \frac{2345}{128}      \zeta_5     \xil
    + \frac{2995}{384}      \zeta_5     \xil^{2}
    + \frac{55}{128}      \zeta_5     \xil^{3}
    - \frac{35}{192}      \zeta_5     \xil^{4}
\right. \Break \left. \qquad \qquad
    + \frac{2065}{128}      \zeta_5
    + \frac{44555}{2304}      \xil
    + \frac{127487}{20736}      \xil^{2}
    - \frac{271}{192}      \xil^{3}
    - \frac{149}{192}      \xil^{4}
    + \frac{3}{32}      \xil^{5}
    - \frac{1}{64}      \xil^{6}
\right]\Break
+ \;     \alvp^{3}     C_A^{2}     C_F\;
\left[
    - 2
    + \frac{43}{32}      \zeta_3     \xil
    + \frac{101}{32}      \zeta_3     \xil^{2}
    + \frac{17}{32}      \zeta_3     \xil^{3}
    - \frac{1497}{32}      \zeta_3
    - \frac{325}{32}      \zeta_5     \xil
    - \frac{215}{32}      \zeta_5     \xil^{2}
\right. \Break \left. \qquad \qquad
    - \frac{35}{32}      \zeta_5     \xil^{3}
    + \frac{1695}{32}      \zeta_5
    - \frac{1}{2}      \xil
    + \frac{1}{4}      \xil^{2}
\right] 
+ \;     \alvp^{3}     C_F     T^{2}     n_f^{2}\;
\left[
    - \frac{7402}{81} 
    + \frac{608}{9}      \zeta_3
\right]\Break
+ \;     \alvp^{3}     C_A^{2}     T     n_f\;
\left[
    + \frac{682607}{2592} 
    + \frac{1889}{72}      \zeta_3     \xil
    - \frac{25}{6}      \zeta_3     \xil^{2}
    + \frac{3733}{72}      \zeta_3
    - \frac{27}{2}      \zeta_4
    - \frac{160}{3}      \zeta_5
    - \frac{521}{96}      \xil
\right. \Break \left. \qquad \qquad
    - \frac{9511}{2592}      \xil^{2}
    - \frac{5}{6}      \xil^{3}
    + \frac{5}{12}      \xil^{4}
\right] 
+ \;     \alvp^{3}     C_F^{2}     T     n_f\;
\left[
    - \frac{286}{9} 
    - \frac{296}{3}      \zeta_3
    + 160     \zeta_5
\right]\Break
+ \;     \alvp^{3}     C_A     C_F     T     n_f\;
\left[
    + \frac{181711}{648} 
    + 8     \zeta_3     \xil^{2}
    - \frac{1438}{9}      \zeta_3
    + 18     \zeta_4
    - 80     \zeta_5
    - \frac{55}{6}      \xil^{2}
\right]\Break
+ \;     \alvp^{3}     C_A     T^{2}     n_f^{2}\;
\left[
    - \frac{3415}{81} 
    - \frac{64}{9}      \zeta_3     \xil
    - \frac{248}{9}      \zeta_3
    + \frac{16}{9}      \xil
    - \frac{100}{81}      \xil^{2}
\right] 
{}
{}.\end{eqnarray}
\begin{eqnarray}
\label{T2renMSb}
T_2^{\smsbar} & = & 
 \alvp C_A\;
\left[
    - \frac{37}{12} 
    + \frac{3}{2}      \xil
    + \frac{1}{4}      \xil^{2}
\right] 
+ \;     \alvp     T     n_f\;
\left[
    + \frac{8}{3} 
\right]\Break
+ \;     \alvp^{2}     C_A^{2}\;
\left[
    - \frac{443}{24} 
    + \frac{1}{2}      \zeta_3     \xil
    - \frac{3}{2}      \zeta_3
    + \frac{31}{48}      \xil
    + \frac{59}{72}      \xil^{2}
    - \frac{11}{16}      \xil^{3}
    - \frac{1}{8}      \xil^{4}
\right]\Break
+ \;     \alvp^{2}     C_A     T     n_f\;
\left[
    + \frac{91}{6} 
    + \frac{5}{2}      \xil
    - \frac{2}{9}      \xil^{2}
\right] 
+ \;     \alvp^{2}     C_F     T     n_f\;
\left[
    + 8
\right]\Break
+ \;     \alvp^{3}     C_A^{3}\;
\left[
    - \frac{2108863}{6912} 
    - \frac{787}{36}      \zeta_3     \xil
    - \frac{1093}{96}      \zeta_3     \xil^{2}
    - \frac{95}{48}      \zeta_3     \xil^{3}
    + \frac{7}{48}      \zeta_3     \xil^{4}
    - \frac{19255}{96}      \zeta_3
\right. \Break \left. \qquad \qquad
    + \frac{3805}{96}      \zeta_5     \xil
    + \frac{845}{48}      \zeta_5     \xil^{2}
    + \frac{355}{96}      \zeta_5     \xil^{3}
    + \frac{35}{192}      \zeta_5     \xil^{4}
    + \frac{46985}{192}      \zeta_5
    + \frac{9869}{2304}      \xil
\right. \Break \left. \qquad \qquad
    + \frac{178183}{20736}      \xil^{2}
    + \frac{7039}{2304}      \xil^{3}
    - \frac{57}{128}      \xil^{4}
    + \frac{5}{32}      \xil^{5}
    + \frac{3}{64}      \xil^{6}
\right]\Break
+ \;     \alvp^{3}     C_A^{2}     C_F\;
\left[
    + \frac{53}{2} 
    + \frac{245}{8}      \zeta_3     \xil
    + \frac{49}{2}      \zeta_3     \xil^{2}
    - \frac{5}{8}      \zeta_3     \xil^{3}
    - \frac{9}{16}      \zeta_3     \xil^{4}
    + \frac{7249}{16}      \zeta_3
    - \frac{125}{2}      \zeta_5     \xil
\right. \Break \left. \qquad \qquad
    - 50     \zeta_5     \xil^{2}
    - 5     \zeta_5     \xil^{3}
    - \frac{1025}{2}      \zeta_5
    - \frac{61}{4}      \xil
    + \frac{3}{4}      \xil^{3}
\right]\Break
+ \;     \alvp^{3}     C_A^{2}     T     n_f\;
\left[
    + \frac{279701}{864} 
    + \frac{5}{18}      \zeta_3     \xil
    + \frac{19}{4}      \zeta_3     \xil^{2}
    + \frac{749}{12}      \zeta_3
    + \frac{10}{3}      \zeta_5     \xil
    - \frac{110}{3}      \zeta_5
    + \frac{197}{12}      \xil
\right. \Break \left. \qquad \qquad
    + \frac{3637}{2592}      \xil^{2}
    - \frac{131}{36}      \xil^{3}
    - \frac{1}{3}      \xil^{4}
\right]\Break
+ \;     \alvp^{3}     C_A     C_F     T     n_f\;
\left[
    + \frac{1105}{6} 
    - 24     \zeta_3     \xil
    - 8     \zeta_3     \xil^{2}
    - 208     \zeta_3
    + \frac{320}{3}      \zeta_5
    + 25     \xil
    + \frac{31}{6}      \xil^{2}
\right]\Break
+ \;     \alvp^{3}     C_A     T^{2}     n_f^{2}\;
\left[
    - \frac{1741}{27} 
    + \frac{32}{9}      \zeta_3     \xil
    - 16     \zeta_3
    - \frac{14}{9}      \xil
    - \frac{140}{81}      \xil^{2}
\right]\Break
+ \;     \alvp^{3}     C_F     T^{2}     n_f^{2}\;
\left[
    - \frac{176}{3} 
    + \frac{128}{3}      \zeta_3
\right] 
+ \;     \alvp^{3}     C_F^{2}     T     n_f\;
\left[
    - 4
\right] 
{}
{}.\end{eqnarray}

   \subsection{The Ghost Gluon Vertex}
\label{app:ccg}

\begin{eqnarray}
\label{GammahrenMSb}
\tilde \Gamma_{\mathrm{h}}^{\smsbar} & = & 
1\;
{}
+ \; \alvp C_A\;
\left[
    + \frac{1}{2}      \xil
\right] 
+ \;     \alvp^{2}     C_A^{2}\;
\left[
    - \frac{9}{16}      \zeta_3     \xil
    + \frac{3}{16}      \zeta_3     \xil^{2}
    + \frac{43}{16}      \xil
    + \frac{7}{16}      \xil^{2}
\right]\Break
+ \;     \alvp^{3}     C_A^{3}\;
\left[
    + \frac{725}{192}      \zeta_3     \xil
    + \frac{267}{64}      \zeta_3     \xil^{2}
    - 2     \zeta_3     \xil^{3}
    - \frac{105}{8}      \zeta_5     \xil
    - \frac{95}{16}      \zeta_5     \xil^{2}
    + \frac{35}{16}      \zeta_5     \xil^{3}
\right. \Break \left. \qquad \qquad
    + \frac{3631}{128}      \xil
    + \frac{2089}{384}      \xil^{2}
    + \frac{17}{24}      \xil^{3}
\right] 
+ \;     \alvp^{3}     C_A^{2}     T     n_f\;
\left[
    + \frac{29}{12}      \zeta_3     \xil
    - \frac{493}{48}      \xil
\right]\Break
+ \;     \alvp^{3}     C_A^{2}     C_F\;
\left[
    - 27     \zeta_3     \xil
    - 6     \zeta_3     \xil^{2}
    + \frac{9}{2}      \zeta_3     \xil^{3}
    + \frac{225}{8}      \zeta_5     \xil
    + \frac{75}{8}      \zeta_5     \xil^{2}
    - \frac{15}{4}      \zeta_5     \xil^{3}
\right] 
{}
{}.\end{eqnarray}
\begin{eqnarray}
\label{GammagrenMSb}
\tilde \Gamma_{\mathrm{g}}^{\smsbar} & = & 
1\;
{}
+ \; \alvp C_A\;
\left[
    + \frac{3}{4} 
    + \frac{1}{4}      \xil
\right]\Break
+ \;     \alvp^{2}     C_A^{2}\;
\left[
    + \frac{599}{96} 
    - \frac{9}{16}      \zeta_3     \xil
    + \frac{3}{16}      \zeta_3     \xil^{2}
    + \frac{97}{32}      \xil
    + \frac{1}{4}      \xil^{2}
\right] 
+ \;     \alvp^{2}     C_A     T     n_f\;
\left[
    - \frac{29}{12} 
\right]\Break
+ \;     \alvp^{3}     C_A^{3}\;
\left[
    + \frac{43273}{432} 
    - \frac{817}{192}      \zeta_3     \xil
    + \frac{293}{64}      \zeta_3     \xil^{2}
    + \frac{3}{64}      \zeta_3     \xil^{3}
    + \frac{783}{64}      \zeta_3
    - \frac{85}{16}      \zeta_5     \xil
    - \frac{325}{64}      \zeta_5     \xil^{2}
\right. \Break \left. \qquad \qquad
    + \frac{5}{16}      \zeta_5     \xil^{3}
    - \frac{875}{64}      \zeta_5
    + \frac{28039}{768}      \xil
    + \frac{4091}{768}      \xil^{2}
    + \frac{31}{48}      \xil^{3}
\right]\Break
+ \;     \alvp^{3}     C_A^{2}     C_F\;
\left[
    + \frac{27}{4} 
    - \frac{249}{16}      \zeta_3     \xil
    - \frac{111}{16}      \zeta_3     \xil^{2}
    + \frac{3}{16}      \zeta_3     \xil^{3}
    - \frac{639}{16}      \zeta_3
    + \frac{165}{8}      \zeta_5     \xil
    + \frac{15}{2}      \zeta_5     \xil^{2}
\right. \Break \left. \qquad \qquad
    + \frac{225}{8}      \zeta_5
    - \frac{21}{4}      \xil
    + \frac{3}{4}      \xil^{2}
\right] 
+ \;     \alvp^{3}     C_A     C_F     T     n_f\;
\left[
    - 16
    + 12     \zeta_3
\right]\Break
+ \;     \alvp^{3}     C_A^{2}     T     n_f\;
\left[
    - \frac{15143}{216} 
    + \frac{55}{24}      \zeta_3     \xil
    - \frac{49}{8}      \zeta_3
    - \frac{357}{32}      \xil
\right] 
+ \;     \alvp^{3}     C_A     T^{2}     n_f^{2}\;
\left[
    + \frac{280}{27} 
\right] 
{}
{}.\end{eqnarray}

   \subsection{The Quark Gluon Vertex}
\label{app:qqg}

\begin{eqnarray}
\label{LambdaqrenMSb}
\Lambda_{\mathrm{q}}^{\smsbar} & = & 
1\;
{}
+ \; \alvp C_A\;
\left[
    + 1
    + \frac{1}{2}      \xil
\right] 
+ \;     \alvp     C_F\;
\left[
    + \xil
\right]\Break
+ \;     \alvp^{2}     C_A^{2}\;
\left[
    + \frac{2015}{192} 
    - \frac{3}{2}      \zeta_3
    + \frac{181}{64}      \xil
    + \frac{13}{16}      \xil^{2}
\right] 
+ \;     \alvp^{2}     C_A     T     n_f\;
\left[
    - \frac{95}{24} 
\right]\Break
+ \;     \alvp^{2}     C_A     C_F\;
\left[
    + \frac{41}{4} 
    - 3     \zeta_3     \xil
    - 3     \zeta_3
    + \frac{15}{2}      \xil
    + \frac{13}{8}      \xil^{2}
\right] 
+ \;     \alvp^{2}     C_F     T     n_f\;
\left[
    - \frac{7}{2} 
\right] 
+ \;     \alvp^{2}     C_F^{2}\;
\left[
    - \frac{5}{8} 
\right]\Break
+ \;     \alvp^{3}     C_A^{3}\;
\left[
    + \frac{2255345}{15552} 
    - \frac{119}{192}      \zeta_3     \xil
    - \frac{19}{32}      \zeta_3     \xil^{2}
    - \frac{37}{64}      \zeta_3     \xil^{3}
    - \frac{1817}{72}      \zeta_3
    + \frac{3}{16}      \zeta_4     \xil
    + \frac{3}{64}      \zeta_4     \xil^{2}
\right. \Break \left. \qquad \qquad
    + \frac{9}{64}      \zeta_4
    - \frac{105}{32}      \zeta_5     \xil
    + \frac{5}{32}      \zeta_5     \xil^{2}
    + \frac{15}{32}      \zeta_5     \xil^{3}
    - \frac{335}{32}      \zeta_5
    + \frac{6511}{192}      \xil
    + \frac{3295}{384}      \xil^{2}
\right. \Break \left. \qquad \qquad
    + \frac{235}{128}      \xil^{3}
\right] 
+ \;     \alvp^{3}     C_A^{2}     T     n_f\;
\left[
    - \frac{169525}{1944} 
    - \frac{1}{6}      \zeta_3     \xil
    - \frac{2}{9}      \zeta_3
    - \frac{9}{2}      \zeta_4
    - \frac{491}{48}      \xil
\right]\Break
+ \;     \alvp^{3}     C_A^{2}     C_F\;
\left[
    + \frac{331393}{1296} 
    - 41     \zeta_3     \xil
    - \frac{45}{8}      \zeta_3     \xil^{2}
    + \frac{5}{12}      \zeta_3     \xil^{3}
    - \frac{3631}{24}      \zeta_3
    + \frac{3}{8}      \zeta_4     \xil
    + \frac{3}{16}      \zeta_4     \xil^{2}
\right. \Break \left. \qquad \qquad
    - \frac{69}{16}      \zeta_4
    + \frac{35}{8}      \zeta_5     \xil
    + \frac{5}{4}      \zeta_5     \xil^{2}
    - \frac{5}{8}      \zeta_5     \xil^{3}
    + \frac{125}{2}      \zeta_5
    + \frac{13117}{144}      \xil
    + \frac{311}{16}      \xil^{2}
    + \frac{11}{3}      \xil^{3}
\right]\Break
+ \;     \alvp^{3}     C_A     C_F^{2}\;
\left[
    - \frac{253}{6} 
    - 17     \zeta_3     \xil
    + \zeta_3     \xil^{3}
    + 44     \zeta_3
    + 6     \zeta_4
    + 20     \zeta_5     \xil
    - 20     \zeta_5
    + \frac{59}{16}      \xil
    + \frac{3}{2}      \xil^{2}
\right. \Break \left. \qquad \qquad
    - \frac{1}{8}      \xil^{3}
\right] 
+ \;     \alvp^{3}     C_A     C_F     T     n_f\;
\left[
    - \frac{121637}{648} 
    + 8     \zeta_3     \xil
    + \frac{118}{3}      \zeta_3
    + 6     \zeta_4
    - \frac{1067}{36}      \xil
\right]\Break
+ \;     \alvp^{3}     C_A     T^{2}     n_f^{2}\;
\left[
    + \frac{5161}{486} 
    + \frac{8}{9}      \zeta_3
\right] 
+ \;     \alvp^{3}     C_F     T^{2}     n_f^{2}\;
\left[
    + \frac{1570}{81} 
\right]\Break
+ \;     \alvp^{3}     C_F^{2}     T     n_f\;
\left[
    - \frac{79}{6} 
    + 16     \zeta_3
    - \frac{3}{2}      \xil
\right] 
+ \;     \alvp^{3}     C_F^{3}\;
\left[
    - \frac{73}{12} 
    - \frac{2}{3}      \zeta_3     \xil^{3}
    + \frac{7}{8}      \xil
\right] 
{}
{}.\end{eqnarray}
\begin{eqnarray}
\label{LambdaqTrenMSb}
\Lambda_{\mathrm{q}}^{T\,\smsbar} & = & 
 \alvp C_A\;
\left[
    + \frac{9}{4} 
    - \xil
    - \frac{1}{4}      \xil^{2}
\right] 
+ \;     \alvp     C_F\;
\left[
    - 2
\right]\Break
+ \;     \alvp^{2}     C_A^{2}\;
\left[
    + \frac{523}{24} 
    - \frac{11}{8}      \zeta_3     \xil
    - \frac{1}{16}      \zeta_3     \xil^{2}
    + \frac{11}{16}      \zeta_3
    - \frac{145}{48}      \xil
    - \frac{215}{144}      \xil^{2}
    - \frac{1}{16}      \xil^{3}
    + \frac{1}{16}      \xil^{4}
\right]\Break
+ \;     \alvp^{2}     C_A     C_F\;
\left[
    - \frac{505}{18} 
    + \frac{13}{4}      \xil
    - \frac{1}{2}      \xil^{2}
    - \frac{1}{4}      \xil^{3}
\right] 
+ \;     \alvp^{2}     C_F     T     n_f\;
\left[
    + \frac{52}{9} 
\right]\Break
+ \;     \alvp^{2}     C_A     T     n_f\;
\left[
    - \frac{16}{3} 
    - 4     \zeta_3
    - \frac{2}{3}      \xil
    - \frac{5}{9}      \xil^{2}
\right] 
+ \;     \alvp^{2}     C_F^{2}\;
\left[
    + 9
    - 2     \xil
\right]\Break
+ \;     \alvp^{3}     C_A^{3}\;
\left[
    + \frac{2844547}{6912} 
    - \frac{2221}{96}      \zeta_3     \xil
    - \frac{203}{192}      \zeta_3     \xil^{2}
    + \frac{89}{48}      \zeta_3     \xil^{3}
    + \frac{29}{192}      \zeta_3     \xil^{4}
    + \frac{565}{9}      \zeta_3
\right. \Break \left. \qquad \qquad
    + \frac{655}{48}      \zeta_5     \xil
    + \frac{385}{96}      \zeta_5     \xil^{2}
    - \frac{35}{48}      \zeta_5     \xil^{3}
    - \frac{35}{192}      \zeta_5     \xil^{4}
    - \frac{23575}{192}      \zeta_5
    - \frac{268897}{6912}      \xil
\right. \Break \left. \qquad \qquad
    - \frac{40489}{2304}      \xil^{2}
    - \frac{10445}{2304}      \xil^{3}
    - \frac{155}{576}      \xil^{4}
    + \frac{1}{16}      \xil^{5}
    - \frac{1}{64}      \xil^{6}
\right]\Break
+ \;     \alvp^{3}     C_A^{2}     C_F\;
\left[
    - \frac{1450057}{2592} 
    + \frac{389}{48}      \zeta_3     \xil
    + \frac{35}{12}      \zeta_3     \xil^{2}
    - \frac{1}{16}      \zeta_3     \xil^{3}
    - \frac{1939}{8}      \zeta_3
    - \frac{265}{24}      \zeta_5     \xil
\right. \Break \left. \qquad \qquad
    + \frac{5}{24}      \zeta_5     \xil^{2}
    + \frac{5}{8}      \zeta_5     \xil^{3}
    + \frac{8665}{24}      \zeta_5
    + \frac{10469}{288}      \xil
    - \frac{299}{96}      \xil^{2}
    - \frac{149}{36}      \xil^{3}
    - \frac{15}{32}      \xil^{4}
    + \frac{1}{16}      \xil^{5}
\right]\Break
+ \;     \alvp^{3}     C_A^{2}     T     n_f\;
\left[
    - \frac{171143}{864} 
    + \frac{275}{18}      \zeta_3     \xil
    - \frac{3}{2}      \zeta_3     \xil^{2}
    - \frac{880}{9}      \zeta_3
    + \frac{20}{3}      \zeta_5
    + \frac{415}{72}      \xil
    - \frac{575}{288}      \xil^{2}
\right. \Break \left. \qquad \qquad
    + \frac{1}{36}      \xil^{3}
    + \frac{5}{18}      \xil^{4}
\right] 
+ \;     \alvp^{3}     C_A     T^{2}     n_f^{2}\;
\left[
    + \frac{700}{27} 
    - \frac{32}{9}      \zeta_3     \xil
    + \frac{112}{9}      \zeta_3
    + \frac{40}{27}      \xil
\right]\Break
+ \;     \alvp^{3}     C_A     C_F     T     n_f\;
\left[
    + \frac{78625}{648} 
    - 4     \zeta_3     \xil
    + 4     \zeta_3     \xil^{2}
    + \frac{196}{3}      \zeta_3
    + \frac{400}{3}      \zeta_5
    - \frac{73}{18}      \xil
    - \frac{113}{24}      \xil^{2}
    - \frac{5}{9}      \xil^{3}
\right]\Break
+ \;     \alvp^{3}     C_A     C_F^{2}\;
\left[
    + \frac{76339}{288} 
    + 6     \zeta_3     \xil
    + 316     \zeta_3
    - \frac{1240}{3}      \zeta_5
    - \frac{3235}{72}      \xil
    - \frac{107}{32}      \xil^{2}
    + \frac{1}{2}      \xil^{3}
\right]\Break
+ \;     \alvp^{3}     C_F     T^{2}     n_f^{2}\;
\left[
    - \frac{2000}{81} 
\right] 
+ \;     \alvp^{3}     C_F^{2}     T     n_f\;
\left[
    + \frac{821}{9} 
    - \frac{160}{3}      \zeta_3
    - 160     \zeta_5
    + \frac{52}{9}      \xil
\right]\Break
+ \;     \alvp^{3}     C_F^{3}\;
\left[
    - \frac{973}{12} 
    - \frac{496}{3}      \zeta_3
    + \frac{640}{3}      \zeta_5
    + 9     \xil
\right] 
{}
{}.\end{eqnarray}
\begin{eqnarray}
\label{LambdagrenMSb}
\Lambda_{\mathrm{g}}^{\smsbar} & = & 
1\;
{}
+ \; \alvp C_A\;
\left[
    + \frac{1}{4} 
    + \frac{3}{4}      \xil
\right] 
+ \;     \alvp     C_F\;
\left[
    - \xil
\right]\Break
+ \;     \alvp^{2}     C_A^{2}\;
\left[
    + \frac{1075}{192} 
    - 3     \zeta_3
    + \frac{181}{64}      \xil
    + \xil^{2}
\right] 
+ \;     \alvp^{2}     C_A     C_F\;
\left[
    + \frac{3}{4} 
    - 3     \zeta_3     \xil
    - 3     \zeta_3
    - \frac{3}{2}      \xil
    - \frac{3}{8}      \xil^{2}
\right]\Break
+ \;     \alvp^{2}     C_A     T     n_f\;
\left[
    - \frac{55}{24} 
\right] 
+ \;     \alvp^{2}     C_F     T     n_f\;
\left[
    + \frac{1}{2} 
\right] 
+ \;     \alvp^{2}     C_F^{2}\;
\left[
    + \frac{19}{8} 
    - 2     \xil^{2}
\right]\Break
+ \;     \alvp^{3}     C_A^{3}\;
\left[
    + \frac{59815}{1944} 
    - \frac{1235}{48}      \zeta_3     \xil
    - \frac{73}{32}      \zeta_3     \xil^{2}
    - \frac{1}{4}      \zeta_3     \xil^{3}
    - \frac{10145}{288}      \zeta_3
    + \frac{3}{16}      \zeta_4     \xil
    + \frac{3}{64}      \zeta_4     \xil^{2}
\right. \Break \left. \qquad \qquad
    + \frac{9}{64}      \zeta_4
    + \frac{35}{4}      \zeta_5     \xil
    - \frac{5}{16}      \zeta_5     \xil^{2}
    + \frac{365}{16}      \zeta_5
    + \frac{5695}{128}      \xil
    + \frac{4033}{384}      \xil^{2}
    + \frac{141}{64}      \xil^{3}
\right]\Break
+ \;     \alvp^{3}     C_A^{2}     C_F\;
\left[
    + \frac{21971}{648} 
    + \frac{55}{2}      \zeta_3     \xil
    - \frac{5}{4}      \zeta_3     \xil^{2}
    + \frac{5}{12}      \zeta_3     \xil^{3}
    - \frac{527}{3}      \zeta_3
    + \frac{3}{8}      \zeta_4     \xil
    + \frac{3}{16}      \zeta_4     \xil^{2}
    - \frac{69}{16}      \zeta_4
\right. \Break \left. \qquad \qquad
    - \frac{35}{2}      \zeta_5     \xil
    + \frac{5}{2}      \zeta_5     \xil^{2}
    + 140     \zeta_5
    - \frac{13117}{288}      \xil
    - \frac{159}{16}      \xil^{2}
    - \frac{173}{96}      \xil^{3}
\right]\Break
+ \;     \alvp^{3}     C_A^{2}     T     n_f\;
\left[
    - \frac{89777}{3888} 
    + \frac{17}{6}      \zeta_3     \xil
    + \frac{359}{18}      \zeta_3
    - \frac{9}{2}      \zeta_4
    - 20     \zeta_5
    - 14     \xil
\right]\Break
+ \;     \alvp^{3}     C_A     C_F^{2}\;
\left[
    + \frac{1259}{32} 
    - 17     \zeta_3     \xil
    + \frac{9}{2}      \zeta_3     \xil^{2}
    + \frac{1}{2}      \zeta_3     \xil^{3}
    + 82     \zeta_3
    + 6     \zeta_4
    + 20     \zeta_5     \xil
    - 100     \zeta_5
\right. \Break \left. \qquad \qquad
    - \frac{747}{32}      \xil
    - \frac{29}{2}      \xil^{2}
    - \frac{23}{8}      \xil^{3}
\right] 
+ \;     \alvp^{3}     C_A     T^{2}     n_f^{2}\;
\left[
    + \frac{1741}{486} 
    + \frac{8}{9}      \zeta_3
\right]\Break
+ \;     \alvp^{3}     C_A     C_F     T     n_f\;
\left[
    - \frac{6823}{324} 
    - 2     \zeta_3     \xil
    + \frac{58}{3}      \zeta_3
    + 6     \zeta_4
    + \frac{761}{72}      \xil
\right] 
+ \;     \alvp^{3}     C_F     T^{2}     n_f^{2}\;
\left[
    - \frac{302}{81} 
\right]\Break
+ \;     \alvp^{3}     C_F^{2}     T     n_f\;
\left[
    - \frac{45}{2} 
    + 16     \zeta_3
    + \frac{19}{2}      \xil
\right] 
+ \;     \alvp^{3}     C_F^{3}\;
\left[
    - \frac{109}{12} 
    - \frac{2}{3}      \zeta_3     \xil^{3}
    + \frac{41}{8}      \xil
\right] 
{}
{}.\end{eqnarray}
\begin{eqnarray}
\label{LambdagTrenMSb}
\Lambda_{\mathrm{g}}^{T\,\smsbar} & = & 
 \alvp C_A\;
\left[
    + \frac{1}{2} 
    - \frac{1}{2}      \xil
\right] 
+ \;     \alvp     C_F\;
\left[
    + 2     \xil
\right] 
+ \;     \alvp^{2}     C_A^{2}\;
\left[
    + \frac{167}{36} 
    - \frac{1}{2}      \zeta_3     \xil
    - \frac{5}{2}      \zeta_3
    - \frac{29}{24}      \xil
    - \frac{5}{8}      \xil^{2}
\right]\Break
+ \;     \alvp^{2}     C_A     C_F\;
\left[
    + \frac{43}{6} 
    + 4     \zeta_3
    + 9     \xil
    + \frac{3}{2}      \xil^{2}
\right] 
+ \;     \alvp^{2}     C_A     T     n_f\;
\left[
    - \frac{7}{9} 
\right] 
+ \;     \alvp^{2}     C_F     T     n_f\;
\left[
    - 4
\right]\Break
+ \;     \alvp^{2}     C_F^{2}\;
\left[
    - 3
    + 2     \xil^{2}
\right] 
+ \;     \alvp^{3}     C_A     C_F     T     n_f\;
\left[
    - \frac{15827}{108} 
    + 8     \zeta_3     \xil
    - \frac{40}{9}      \zeta_3
    - \frac{112}{3}      \xil
\right]\Break
+ \;     \alvp^{3}     C_A^{3}\;
\left[
    + \frac{1017553}{10368} 
    + \frac{857}{72}      \zeta_3     \xil
    + \frac{7}{16}      \zeta_3     \xil^{2}
    - \frac{7373}{144}      \zeta_3
    - \frac{115}{12}      \zeta_5     \xil
    - \frac{205}{12}      \zeta_5
    - \frac{36727}{1728}      \xil
\right. \Break \left. \qquad \qquad
    - \frac{2309}{384}      \xil^{2}
    - \frac{43}{32}      \xil^{3}
\right]\Break
+ \;     \alvp^{3}     C_A^{2}     C_F\;
\left[
    + \frac{39011}{216} 
    - 63     \zeta_3     \xil
    - \frac{9}{4}      \zeta_3     \xil^{2}
    + \frac{3917}{36}      \zeta_3
    + 25     \zeta_5     \xil
    - \frac{305}{3}      \zeta_5
    + \frac{4091}{32}      \xil
\right. \Break \left. \qquad \qquad
    + \frac{2263}{96}      \xil^{2}
    + \frac{63}{16}      \xil^{3}
\right]\Break
+ \;     \alvp^{3}     C_A^{2}     T     n_f\;
\left[
    - \frac{61085}{1296} 
    - \frac{2}{9}      \zeta_3     \xil
    - \frac{14}{9}      \zeta_3
    + \frac{80}{3}      \zeta_5
    + \frac{3197}{432}      \xil
\right] 
+ \;     \alvp^{3}     C_F^{3}\;
\left[
    + 3
    - \frac{17}{4}      \xil
\right]\Break
+ \;     \alvp^{3}     C_A     C_F^{2}\;
\left[
    - \frac{3515}{48} 
    + 6     \zeta_3     \xil
    - 6     \zeta_3     \xil^{2}
    - \frac{184}{3}      \zeta_3
    + \frac{280}{3}      \zeta_5
    + \frac{361}{16}      \xil
    + 17     \xil^{2}
    + \frac{11}{4}      \xil^{3}
\right]\Break
+ \;     \alvp^{3}     C_A     T^{2}     n_f^{2}\;
\left[
    + \frac{260}{81} 
\right] 
+ \;     \alvp^{3}     C_F     T^{2}     n_f^{2}\;
\left[
    + \frac{208}{9} 
\right] 
+ \;     \alvp^{3}     C_F^{2}     T     n_f\;
\left[
    + \frac{28}{3} 
    - 11     \xil
\right] 
{}
{}.\end{eqnarray}

\section{Renormalization Group Coefficients}
\label{app:anomdims}

\subsection{The $\beta$-function}

\begin{eqnarray}
\label{beta3l}
\beta & = & 
 \alvp^{2}\;
\left[
    - \frac{11}{3}      C_A
    + \frac{4}{3}      T     n_f
\right] 
+ \;     \alvp^{3}\;
\left[
    + \frac{20}{3}      C_A     T     n_f
    - \frac{34}{3}      C_A^{2}
    + 4     C_F     T     n_f
\right]\Break
+ \;     \alvp^{4}\;
\left[
    + \frac{205}{9}      C_A     C_F     T     n_f
    - \frac{158}{27}      C_A     T^{2}     n_f^{2}
    + \frac{1415}{27}      C_A^{2}     T     n_f
    - \frac{2857}{54}      C_A^{3}
\right. \Break \left. \qquad \qquad
    - \frac{44}{9}      C_F     T^{2}     n_f^{2}
    - 2     C_F^{2}     T     n_f
\right] 
{}
{}.\end{eqnarray}

\subsection{The Gluon Field Anomalous Dimension}

\begin{eqnarray}
\label{gammag}
\gamma_3 & = & 
 \alvp C_A\;
\left[
    + \frac{13}{6} 
    - \frac{1}{2}      \xil
\right] 
+ \;     \alvp     n_f     T\;
\left[
    - \frac{4}{3} 
\right]\Break
+ \;     \alvp^{2}     C_A^{2}\;
\left[
    + \frac{59}{8} 
    - \frac{11}{8}      \xil
    - \frac{1}{4}      \xil^{2}
\right] 
+ \;     \alvp^{2}     C_A     n_f     T\;
\left[
    - 5
\right] 
+ \;     \alvp^{2}     C_F     n_f     T\;
\left[
    - 4
\right]\Break
+ \;     \alvp^{3}     C_A^{3}\;
\left[
    + \frac{9965}{288} 
    - \frac{3}{4}      \zeta_3     \xil
    - \frac{3}{16}      \zeta_3     \xil^{2}
    - \frac{9}{16}      \zeta_3
    - \frac{167}{32}      \xil
    - \frac{33}{32}      \xil^{2}
    - \frac{7}{32}      \xil^{3}
\right]\Break
+ \;     \alvp^{3}     C_A^{2}     n_f     T\;
\left[
    - \frac{911}{18} 
    + 18     \zeta_3
    + 2     \xil
\right] 
+ \;     \alvp^{3}     C_A     C_F     n_f     T\;
\left[
    - \frac{5}{18} 
    - 24     \zeta_3
\right]\Break
+ \;     \alvp^{3}     C_A     n_f^{2}     T^{2}\;
\left[
    + \frac{76}{9} 
\right] 
+ \;     \alvp^{3}     C_F     n_f^{2}     T^{2}\;
\left[
    + \frac{44}{9} 
\right] 
+ \;     \alvp^{3}     C_F^{2}     n_f     T\;
\left[
    + 2
\right] 
{}
{}.\end{eqnarray}

\subsection{The Ghost Field Anomalous Dimension}

\begin{eqnarray}
\label{gammah}
\tilde \gamma_3 & = & 
 \alvp C_A\;
\left[
    + \frac{3}{4} 
    - \frac{1}{4}      \xil
\right] 
+ \;     \alvp^{2}     C_A^{2}\;
\left[
    + \frac{95}{48} 
    + \frac{1}{16}      \xil
\right] 
+ \;     \alvp^{2}     C_A     n_f     T\;
\left[
    - \frac{5}{6} 
\right]\Break
+ \;     \alvp^{3}     C_A     C_F     n_f     T\;
\left[
    - \frac{45}{4} 
    + 12     \zeta_3
\right] 
+ \;     \alvp^{3}     C_A     n_f^{2}     T^{2}\;
\left[
    - \frac{35}{27} 
\right] 
+ \;     \alvp^{3}     C_A^{2}     n_f     T\;
\left[
    - \frac{97}{108} 
    - 9     \zeta_3
    + \frac{7}{8}      \xil
\right]\Break
+ \;     \alvp^{3}     C_A^{3}\;
\left[
    + \frac{15817}{1728} 
    + \frac{3}{8}      \zeta_3     \xil
    + \frac{3}{32}      \zeta_3     \xil^{2}
    + \frac{9}{32}      \zeta_3
    - \frac{17}{32}      \xil
    - \frac{3}{32}      \xil^{2}
    - \frac{3}{64}      \xil^{3}
\right] 
{}
{}.\end{eqnarray}

\subsection{The Quark Field Anomalous Dimension}

\begin{eqnarray}
\label{gammaq}
\gamma_2 & = & 
 \alvp C_F\;
\left[
    - \xil
\right] 
+ \;     \alvp^{2}     C_A     C_F\;
\left[
    - \frac{25}{4} 
    - 2     \xil
    - \frac{1}{4}      \xil^{2}
\right] 
+ \;     \alvp^{2}     C_F     n_f     T\;
\left[
    + 2
\right] 
+ \;     \alvp^{2}     C_F^{2}\;
\left[
    + \frac{3}{2} 
\right]\Break
+ \;     \alvp^{3}     C_A^{2}     C_F\;
\left[
    - \frac{9155}{144} 
    - \frac{3}{4}      \zeta_3     \xil
    - \frac{3}{8}      \zeta_3     \xil^{2}
    + \frac{69}{8}      \zeta_3
    - \frac{263}{32}      \xil
    - \frac{39}{32}      \xil^{2}
    - \frac{5}{16}      \xil^{3}
\right]\Break
+ \;     \alvp^{3}     C_A     C_F     n_f     T\;
\left[
    + \frac{287}{9} 
    + \frac{17}{4}      \xil
\right] 
+ \;     \alvp^{3}     C_A     C_F^{2}\;
\left[
    + \frac{143}{4} 
    - 12     \zeta_3
\right] 
+ \;     \alvp^{3}     C_F     n_f^{2}     T^{2}\;
\left[
    - \frac{20}{9} 
\right]\Break
+ \;     \alvp^{3}     C_F^{2}     n_f     T\;
\left[
    - 3
\right] 
+ \;     \alvp^{3}     C_F^{3}\;
\left[
    - \frac{3}{2} 
\right] 
{}
{}.\end{eqnarray}

\end{appendix}

\end{document}